\documentclass[aps,prb,twocolumn,showpacs,superscriptaddress]{revtex4-2}
\usepackage{graphicx}
\usepackage{amsmath}
\usepackage{multirow}
\usepackage{tabularx}
\usepackage{color}
\usepackage{xcolor}
\usepackage{ulem}
\usepackage{comment}
\usepackage[colorlinks,linkcolor=blue,urlcolor=blue,anchorcolor=blue,citecolor=blue]{hyperref}

\begin{document}
\preprint{APS/123-QED}
\title{Pairing symmetry and superconductivity in La$_3$Ni$_2$O$_7$ thin films}
% \date{\today}

\author{Wenyuan Qiu}
\thanks{These authors contributed equally to this work}
\author{Zhihui Luo}
\thanks{These authors contributed equally to this work}
\affiliation{Institute of Neutron Science and Technology,  State Key Laboratory of Optoelectronic Materials and Technologies, School of Physics, Sun Yat-Sen University, Guangzhou, 510275, China}
\author{Xunwu Hu}
\affiliation{Department of Physics, College of Physics and Optoelectronic Engineering, Jinan University, Guangzhou, 510632, China}
\author{Dao-Xin Yao}
\email{yaodaox@mail.sysu.edu.cn}
\affiliation{Institute of Neutron Science and Technology,  State Key Laboratory of Optoelectronic Materials and Technologies, School of Physics, Sun Yat-Sen University, Guangzhou, 510275, China}

\begin{abstract}
The recent discovery of superconductivity with a transition temperature $T_c$\ over 40 K in La$_3$Ni$_2$O$_7$ and (La,Pr)$_{3}$Ni$_2$O$_7$ thin films at ambient pressure marks an important step in the field of nickelate superconductors. 
Here, we perform a renormalized mean-field theory study of the superconductivity in $\mathrm{La_3Ni_2O_7}$ thin films, using a bilayer two-orbital $t-J$ model.
Our result reveals an $s_\pm$-wave pairing symmetry driven by the strong interlayer superexchange coupling of $d_{z^2}$ orbital, resembling the pressurized bulk case.
Also, we roughly reproduce the experimentally observed nodeless shape of the superconducting gap at the $\beta$ pocket and the superconducting $T_c$.
In addition, by analysing the orbital-resolved pairing configurations and their projections onto Fermi surface, we find that the nodeless feature of $\beta$ pocket is related to the interlayer pairing within both $d_{z^2}$ and $d_{x^2-y^2}$ orbitals. 
Moreover, we identify a formation of the inplane inter-orbital $d$-wave pairing between $d_{z^2}$ and $d_{x^2-y^2}$ orbitals, which can even enhance the dominated interlayer $s_\pm$-wave.
Our study particularly highlights the diverse relations of different pairing channels in $\mathrm{La_3Ni_2O_7}$ that holds a complex Fermi surface.
% {\color{red} Our study particularly highlights the diverse relations of different pairing channels in $\mathrm{La_3Ni_2O_7}$ that holds a complex Fermi surface.}
%This is a unique feature for $\mathrm{La_3Ni_2O_7}$ lattice where both pairings fall in the $A_{1g}$ channel.
\end{abstract}

\maketitle

\section{\label{sec:1}Introduction}
% {\it Introduction.}$-$
The discovery of superconductivity with a transition temperature $T_c$ near 80 K in the Ruddlesden-Popper (RP) bilayer nickelate La$_3$Ni$_2$O$_7$ (LNO) ~\cite{Sunsignatures2023} has gained widespread interests in the field of high-$T_c$ superconductivity,
and it was soon followed by another observation of superconductivity in trilayer nickelate La$_4$Ni$_3$O$_{10}$  with $T_c$=20-30 K~\cite{zhu_superconductivity_2024,li_signature_2024}, both  under high pressure.
The discovery of RP nickelate-based superconductors has marked a breakthrough in the field of high-$T_c$ superconductivity, and therefore raised considerable subsequent investigations both  theoretically~\cite{luo2023bilayer, zhang2023electronic,lechermann2023electronic,luo2024high, wu2024superexchange,  cpl_41_7_077402,   shilenko2023correlated, yang2023interlayer, zhang2023trends,christiansson2023correlated,shen2023effective, oh2023type, liu2023s,liao2023electron,yang2023possible, PhysRevB.110.014503, PhysRevB.111.075140,PhysRevB.110.235155,ouyang2024absence,PhysRevB.109.144511, heier2024competing, zhang2024structural, zhang2024electronic,tian2024correlation, ryee2024quenched,zhang2024strong, ni_spin_2025, lu2024interlayer, qu2024bilayer, yang2024strong, fan2024superconductivity,   sakakibara2024possible, cao2024flat, jiang2024pressure,chen2025charge} and experimentally~\cite{yang_orbital-dependent_2024, NPzhang,Hou_2023, PhysRevB.110.134520, wang_pressure-induced_2024, liu2024electronic, li_signature_2024, Li_2025}. 
However, the need for high pressure has largely confined some critical experimental techniques for a direct probe of the superconducting phase, such as angle resolved photoemission spectroscopy (ARPES) and scanning tunneling microscopy (STM), which prevents a decisive confirmation of the superconducting pairing among various theoretical proposals~\cite{liu2023s,fan2024superconductivity}.
On the other hand,  achieving ambient-pressure superconductivity  is a crucial step for practical considerations, and in this respect, there are some suggestions on the possible route, such as using chemical substitute~\cite{pan_effect_2024, zhang_trends_2023, li2025ambientpressuregrowthbilayer,qiu2025interlayer} or applying strain~\cite{Osada_2025, wang2025electronicstructurecompressivelystrained, bhatt2025resolvingstructuraloriginssuperconductivity, liu2025superconductivitynormalstatetransportcompressively}.

Very recently, superconductivity with  $T_c$ over 40 K in LNO~\cite{ko_signatures_2025} and (La,Pr)$_{3}$Ni$_2$O$_7$~\cite{zhou_ambient-pressure_2025} thin films at ambient pressure have been reported.
%which is a important breakthrough in the field of nickelate superconductors. 
%Various experiments have been conducted on these nickelate thin films to explore the mechanism of superconductivity behind them.
The X-ray absorption spectroscopy (XAS) analysis and scanning transmission electron microscopy (STEM) reveal that, 
the Ni ions in LNO thin films maintain a mixed valence state and the apical Ni-O-Ni bond angle approaches 180$^\circ$~\cite{ko_signatures_2025},  which closely resembles the high-pressure bulk phase of LNO.
The ARPES measurement~\cite{li2025photoemissionevidencemultiorbitalholedoping} in thin films also confirms the dominant  Ni-$d_{x^2-y^2}$ and $d_{z^2}$ orbital spectral weights around the Fermi energy, which is consistent with DFT calculations~\cite{hu_electronic_2025, yue2025correlatedelectronicstructuresunconventional, shi2025effectcarrierdopingthickness, liu2025superconductivitynormalstatetransportcompressively, bhatt2025resolvingstructuraloriginssuperconductivity}.
Moreover, a direct ARPES measurement of the superconducting electronic structure on the Fermi surface (FS) was  performed recently~\cite{shen2025anomalousenergygapsuperconducting},
which reveals a notable gap opening on the $\beta$ FS sheet and without showing any node along the Brillouin zone (BZ) diagonal direction. 
This observation is accompanied by another STM measurement on superconducting state~\cite{fan2025superconductinggapsrevealedstm}, which shows a two-gap structure on the FS and a fitting to such gaps demonstrates a preferable anisotropic $s$-wave pairing.
The two results are generally prone to the idea of the predominant $s_\pm$-wave pairing symmetry in the system as predicted by several theoretical studies~\cite{ yang2023possible, liu2023s, luo2024high, sakakibara2024possible, yue2025correlatedelectronicstructuresunconventional}.
%We note one of the 

%The experiment results appear to favor $s_{\pm}$ pairing symmetry~\cite{fan2025superconductinggapsrevealedstm}, and Random phase approximation (RPA) calculation also support $s_{\pm}$ pairing symmetry~\cite{yue2025correlatedelectronicstructuresunconventional}.

\begin{figure}[hb]
	\centering
	\includegraphics[width=1.\linewidth]{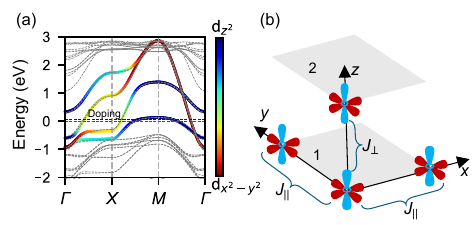}
	\caption{(a) Band structure of the 4-orbital tight-binding model adopted from the DFT simulation of the double-stacked $\mathrm{La_3Ni_2O_7}$ thin film~\cite{hu_electronic_2025}. The red dashed line denotes the undoped case from DFT while the blue one denotes the slightly electron doped case as is used in our study.   (b) Schematic of $d_{x^2-y^2}$ (red) and $d_{z^2}$ (blue) orbitals of the model. Two major superexchange couplings $J_\bot$, $J_{||}$ are drawn. See also the main text.}
	\label{fig:1}
\end{figure}

In this paper, 
%we aim to understand the miscrospic pairing structure in $\mathrm{La_3Ni_2O_7}$ based on existing experiments.
%in Refs.~\cite{shen2025anomalousenergygapsuperconducting,fan2025superconductinggapsrevealedstm}.
using a bilayer $t-J$ model that captures the key ingredient of LNO thin films, we perform a renormalized mean-field theory (RMFT)~\cite{luo2024high} calculation of the superconductivity.
Our result reveals the $s_\pm$-wave pairing and reproduces the experimentally observed nodeless shape of the superconducting gap at $\beta$ pocket in  Ref.~\cite{shen2025anomalousenergygapsuperconducting}.
To gain insight, we also analyse the orbital-resolved pairing bonds and their projections onto Fermi surface. 
Our study highlights the diverse relations of different pairing channels in LNO that holds a complex Fermi surface.

\section{\label{sec:2}Model and method}
% {\it Model and method.}$-$
The bilayer two-orbital $t-J$ model is defined as
\begin{align}
    \mathcal{H} =& \mathcal{H}_t +\mathcal{H}_J, \\
    \mathcal{H}_t =& \sum_{ij}\sum_{\mu\nu}\sum_{\sigma}t_{ij}^{\mu\nu}c_{i\mu\sigma}^{\dagger}c_{j\nu\sigma}-\mu\sum_{i\mu\sigma}n_{i\mu\sigma}\\
    \mathcal{H}_{J} =& 
        J_{\bot}\sum_{i}\boldsymbol{S}_{iz_1}\cdot\boldsymbol{S}_{iz_2} + J_{||}\sum_{\langle ij \rangle}^{\mu=x_1,x_2}\boldsymbol{S}_{i\mu}\cdot\boldsymbol{S}_{j\mu} \\
        +&J_{xz}\sum_{\langle ij \rangle}^{\mu\nu=x_1z_1,x_2z_2}\boldsymbol{S}_{i\mu}\cdot \boldsymbol{S}_{j\nu} 
        -J_H\sum_{i}^{\mu\nu=x_1z_1,x_2z_2}\boldsymbol{S}_{i\mu}\cdot \boldsymbol{S}_{i\nu}. \notag
\end{align}

In this model, the basis is $\Psi=(d_{z_1},d_{x_1},d_{z_2},d_{x_2})^T$ that contains $d_{z^2}$ and $d_{x^2-y^2}$ orbitals for the first and second layers, as shown in Fig.~\ref{fig:1}(b).
$\mathcal{H}_t$ is the tight-binding Hamiltonian adopted from DFT simulation for the double-stacked LNO thin films~\cite{hu_electronic_2025}.
Here, $ij,\mu\nu,\sigma$ denote the lattice, orbital, and spin indices, respectively.
The band structure is shown in Fig.~\ref{fig:1}(a), in which a small electron doping level with chemical potential $\mu=0.05$ is used in our calculation to better agree with the experiments, which gives the electron filling $n_z$=0.8, $n_x$=0.58.
We found that by slightly adjusting the filling, the resulting $\gamma$ pocket around the $M$ point becomes noticeably closer to the ARPES observations~\cite{Liangle2025}.
The observation of Sr diffusion from substrates might explain the differences of electron filling between experiments and theories~\cite{Litheoretical2025}.
For the Heisenberg term $\mathcal{H}_J$, 
$J_\bot$ is the interlayer superexchange coupling of $d_{z^2}$ orbitals, and $J_{||},J_{xz}$ are the inplane nearest neighbor superexchange coupling of $d_{x^2-y^2}$ orbital, and superexchange coupling between $d_{x^2-y^2}$ and $d_{z^2}$, respectively.
Also, the Hund's coupling $J_H$ is included for generality.
To quantitatively estimate $J_{\bot},J_{||},J_{xz}$, we perform the exact diagonalization (ED) calculation and obtain $J_\bot$=0.135\ eV, $J_{||}$=0.084\ eV,
$J_{xz}$=0.03\ eV.
Taking $J_\bot$ as an example,  ED is performed for a 5-site chain containing two $d_{z^2}$ and three $p_z$ orbitals along the $z$ axis, with $p-d$ hoppings obtained from  DFT~\cite{hu_electronic_2025}. Under the atomic  subspace that corresponds to half-filling of $d_{z^2}$ orbital  ($N$=8, $S_z$=0), $J_\bot$ is determined as the energy difference of the lowest singlet and triplet states~\cite{wu2024superexchange}.
Similar method is applied in Refs.~\cite{zhou2025originlocalmagneticexchange,JM2025}.
% {\color{red} Similar method is applied in Refs.~\cite{zhou2025originlocalmagneticexchange,JM2025}.}
 Besides, a realistic $J_H$=1\ eV is also used in this study.

%$i/j$ , $\alpha/\beta$ and $\sigma$ denote site, orbital($d_{z^2}$ or $d_{x^2-y^2}$) and spin, respectively. 
%The hopping parameters $t_{ij}^{\alpha\beta}$ come from the double-stacked two-orbital tight binding model, which we refer to it as One-UC model~\cite{hu_electronic_2025}.
%For the One-UC model, we average the inplane parameters of layer A and layer B and neglect the interstack hoppings. The resulting bands are shown in Fig.~\ref{fig:1} (a). In $H_J$, $J_{\bot}^z$, $J_{||}$ and $J_H$ represent interlayer $d_{z^2}$, inplane $d_{x^2-y^2}$ spin interaction and Hund's interaction, respectively. 
%The $J_{\bot}^z$ can be estimated through exact diagnalization.
%The symbol $z_1/x_1$ ($z_2/x_2$) represents orbital $d_{z^2}/d_{x^2-y^2}$ in Layer A(B).

The above $t-J$ model is solved under the RMFT. The RMFT~\cite{zhang_renormalised_1988} follows a conventional mean-field decomposition of spin exchange, which yields the order parameters $\chi_{ij}^{\mu\nu}=\langle c^{\dagger}_{i\mu\uparrow} c_{j\nu\uparrow}\rangle$ and $\Delta^{\mu\nu}_{ij} = \langle c^{\dagger}_{i\mu\uparrow} c^{\dagger}_{j\nu\downarrow} \rangle$ under the spin symmetry.
%{\color{red} Note that in our model there are 3 types of pairing order parameter, which we simpily denote as $\Delta_x$,$\Delta_y$,$\Delta_z$ as shown in Fig.~\ref{fig:1}b. Here $\Delta_x$, $\Delta_y$ relate $J_{||}$ and $\Delta_z$ relates $J_{\bot}$.}
Besides, $\mathcal{H}_t$ and $\mathcal{H}_J$ are respectively coupled by the Gutzwiller renormalization factors $g_t$ and $g_J$ to recover the strong renormalization from correlation~\cite{gutz}. 
The obtained quadratic Hamiltonian is then calculated self-consistently, during which the electron densities are fixed to certain level.
The eventual pairing symmetry can be determined from the phase structure of various pairing bonds $\Delta^{\mu\nu}_{ij}$.
For the inplane pairing, this is done by allowing $\Delta$  along x and y directions to evolve independently, so a final $s_\pm,d$ wave or $s+id$ wave is possible depending on the superexchanges.
% {\color{red} For the inplane pairing, this is done by allowing $\Delta$  along x and y directions to evolve independently, so a final $s_\pm,d$ wave or $s+id$ wave is possible depending on the superexchanges. }
We refer reader to Ref.~\cite{luo2024high} for details of the method.

\begin{figure}[t]
	\centering
	\includegraphics[width=1.0\linewidth]{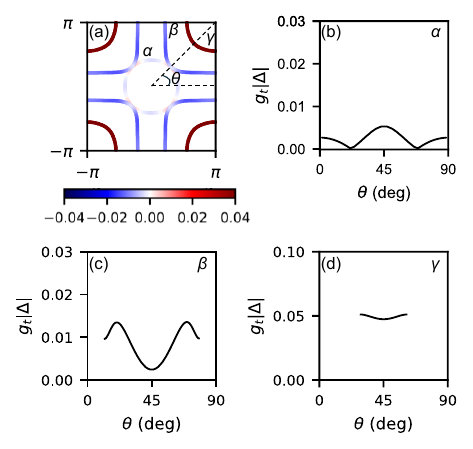}
	\caption{(a) Superconducting gap $g_t\Delta$ projected onto FS for $\mathrm{La_3Ni_2O_7}$ thin films. (b-d) $g_t|\Delta|$ as a function of angle $\theta$ for $\alpha$, $\beta$ and $\gamma$ pockets, respectively. $\theta$ is indicated in (a). }
	\label{fig:gap}
\end{figure}

\section{\label{sec:3}Pairing symmetry and superconducting gap}
% {\it Pairing symmetry and superconducting gap.}$-$
Figure.~\ref{fig:gap}(a) presents the superconducting gap matrix $[g_t\Delta]_{\mu\nu}$ projected onto FS, which demonstrates that $\beta,\gamma$ pockets are nodeless and of opposite sign, while $\alpha$ pocket shows a sign flip.
To better illustrate the gap features, Fig.~\ref{fig:gap}(b-d) show the absolute value $g_t|\Delta|$ as a function of angle $\theta$ for $\alpha,\beta$ and $\gamma$ pockets, respectively.
For $\alpha$ pocket in Fig.~\ref{fig:gap}(b), the gap demonstrates a node between $\theta=45^\circ\sim90^\circ$.
For $\beta$ pocket in Fig.~\ref{fig:gap}(c), the gap shows a parabolic shape centering at $\theta=45^\circ$, which gets slightly decrease when approaching the corner.
Note that the gap shape as well as its magnitude around the BZ diagonal direction from our result are generally coordinated with the ARPES result~\cite{shen2025anomalousenergygapsuperconducting}, which suggests that our RMFT captures the key ingredient in the LNO thin films.
For $\gamma$ pocket in Fig.~\ref{fig:gap}(d), we find the gap magnitude is about 5 times larger than that in Fig.~\ref{fig:gap}(b), with its minimum appearing at the BZ diagonal direction.
Apparently, such large gap opening at $\gamma$ pocket should be attributed to the strong interlayer superexchange coupling $J_{\bot}$, which is also consistent with previous calculations for bulk LNO~\cite{yang2023possible,liu2023s,zhang2024structural,luo2024high}.
Overall, the pairing structure in Fig.~\ref{fig:gap} should be described as the $s_\pm$-wave, which can also be seen from our result that besides the formation of the large interlayer pairing bond for $d_{z^2}$ orbital ($\Delta_{\bot}^z$), the inplane pairing bonds have the same magnitude when along $x,y$ directions for both $d_{x^2-y^2}$ and $d_{z^2}$ orbitals ($\Delta_{||}^x,\Delta_{||}^z)$.
Note that, in our $t-J$ model only the pairing bonds $\Delta_{\bot}^z,\Delta_{||}^x,\Delta_{||}^{xz}$ are associated with the applied $J_\bot,J_{||},J_{xz}$, respectively, while other terms like the interlayer pairing of $d_{x^2-y^2}$ orbital ($\Delta_{\bot}^x$) do not have the corresponding $J$ and should not appear in the calculation.
However, they can still gain finite values as $\Delta_{ij}^{\mu\nu}=\langle n{\rm k}|c_{i\mu\uparrow}^\dagger c^\dagger_{j\nu\downarrow}|n{\rm k}\rangle$, which reflects the pairing tendency towards a global coherent state  at the mean-field level.
In the following we will dive into the role of different pairing bonds to see how they interplay and collectively give rise to the result in Fig.~\ref{fig:gap}.
% {\color{red}However, they can still gain finite values as $\Delta_{ij}^{\mu\nu}=\langle n{\rm k}|c_{i\mu\uparrow}^\dagger c^\dagger_{j\nu\downarrow}|n{\rm k}\rangle$, which reflects the pairing tendency towards a global coherent state  at the mean-field level.
% In the following we will dive into the role of different pairing bonds to see how they interplay and collectively give rise to the result in Fig.~\ref{fig:gap}.}

%For this reason, in Fig.~\ref{fig:gap} we have incorporated all the relevant $\Delta$, and in the following we will illustrate the role of different $\Delta$ on FS.
%and demonstrate that $\Delta_{\bot}^x$ is important from the nodeless feature of $\beta$ pocket at the BZ diagonal.

\section{\label{sec:4}Orbital-resolved superconducting gap}
\begin{figure}[t]
	\centering
	\includegraphics[width=1.0\linewidth]{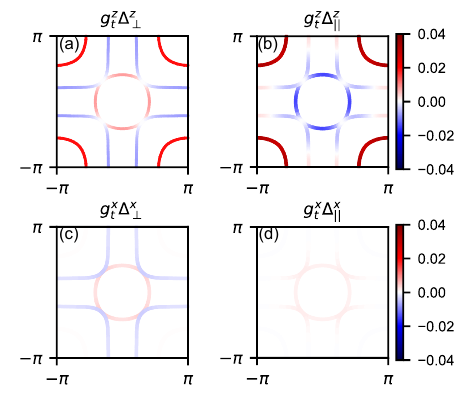}
	\caption{Orbital-resolved superconducting energy gap on FS for (a) $\Delta_{\bot}^z$, (b) $\Delta_{||}^z$, (c) $\Delta_{\bot}^x$, and (d) $\Delta_{||}^x$. $g_t^{z/x}$ is the renormalization factor for $d_{z^2}/d_{x^2-y^2}$ orbital. $\Delta_{\bot/||}^{z/x}$ denotes the interlayer/inplane pairing bond for $d_{z^2}/d_{x^2-y^2}$ orbital.}
	\label{fig:ob_FS}
\end{figure}

% {\it Orbital-resolved superconducting gap.}$-$
Figure.~\ref{fig:ob_FS} presents the orbital-resolved superconducting gap for (a) $\Delta_{\bot}^z$, (b) $\Delta_{||}^z$, (c) $\Delta_{\bot}^x$, and (d) $\Delta_{||}^x$, respectively.  
For $\Delta_{\bot}^z$ in Fig.~\ref{fig:ob_FS}(a), the gap is positive for the bonding branch of $\alpha,\gamma$ pockets, and is negative for the anti-bonding branch of $\beta$ pocket. 
For $\Delta_{||}^z$ in Fig.~\ref{fig:ob_FS}(b), the gap structure clearly demonstrates the sign-flip $s_{\pm}$-wave with the node appearing at the extended BZ boundary (diamond shape).
Also, another node at the BZ diagonal direction in both Fig.~\ref{fig:ob_FS}(a,b) is identified due to the vanishing of hybridization.
%between $d_{x^2-y^2}$ and $d_{z^2}$ orbitals.
For $\Delta_{\bot}^x,\Delta_{||}^x$ in Fig.~\ref{fig:ob_FS}(c,d), similar behavior is observed as compared to Fig.~\ref{fig:ob_FS}(a,b), but with a much smaller magnitude.
It is worth noting that the phases in Fig.~\ref{fig:ob_FS}(a-d) all can be understood in terms of maximizing the total gap magnitude.
Because the large $J_\bot$ will guarantees $\Delta_\bot^z$ to be the leading pairing, other pairings in Fig.~\ref{fig:ob_FS}(c-d) will have to fit into it.
% {\color{red} Because the large $J_\bot$ will guarantees $\Delta_\bot^z$ to be the leading pairing, other pairings in Fig.~\ref{fig:ob_FS}(c-d) will have to fit into it.}
First, Fig.~\ref{fig:ob_FS}(c) will choose the same sign that matches Fig.~\ref{fig:ob_FS}(a) ($sgn\Delta_\bot^z$=$sgn\Delta_\bot^x$). 
This does not conflict with their spatial configuration as the two sides are both connected by $J_H$.
% {\color{red} This does not conflict with their spatial configuration as the two sides are both connected by $J_H$.}
This result clearly explains the experimentally observed nodeless shape of $\beta$ pocket as a joint effort of interlayer pairing for both $d_{x^2-y^2}$ and $d_{z^2}$ orbitals.
Second, due to the inplane $\cos k_x+\cos k_y$ form factor in Fig.~\ref{fig:ob_FS}(b,d),
cancelation of the gap in certain region of FS is unavoidable, which inherently reflects the ``pairing frustration".
In this case $\Delta_{||}^z$ will choose to maximize the gap for $\gamma$ pocket at the price of minimizing the gap for $\alpha$ pocket, while $\Delta_{||}^x$ will no doubt choose to  maximize the gap for $\alpha$ pocket, with slight cancelation at $\beta$ pocket.
In this sense, ``pairing frustration'' is more likely to occur in system with multi-orbital feature and complex Fermi surface where there can not find a single pairing symmetry to perfectly fit in with. More insights into it and the ``gap maximization'' via free energy functional is provided in the Appendix
% {\color{red} In this sense, ``pairing frustration'' is more likely to occur in system with multi-orbital feature and complex Fermi surface where there can not find a single pairing symmetry to perfectly fit in with. More insights into it and the ``gap maximization'' via free energy functional is provided in the Appendix}
In LNO, the obtained gap relation is $sgn\Delta_{\bot}^z=sgn\Delta_\bot^x=sgn\Delta_{||}^x=-sgn\Delta_{||}^z$.
General agreements can be found in Refs.~\cite{yang2023possible,lu2024interlayer}.
%{\color{red}Notably, the sign in Fig.~\ref{fig:ob_FS}(d) is opposite to that in Fig.~\ref{fig:ob_FS}(b), namely, $sgn(\Delta_{||}^x)=-sgn(\Delta_{||}^z)$. 
%This feature is also seen in a slave-boson mean-field study~\cite{lu2024interlayer}, and can be explained as a result of the competition between $J_{\bot}$ preferred $s_\pm$-wave and $J_{||}$ preferred $d$-wave pairings~\cite{sm}.}
%the gap in (d) is opposite to that in (b), i.e., $sgn()$=$-sgn(\Delta_{||}^z)$, 
%which occurs to gain energy from the competition between $J_\bot$ and $J_{||}$.
%In fact, we find that $J_{||}$ always perfers the $d$-wave result at the RMFT mean-field level by setting $J_\bot$=0. Hence, the introduction of the large $J_\bot$ alters pairing at $d_{x^2-y^2}$ orbital to be also the $s_\pm$-wave through hybridization.
%Finally, combining Fig.~\ref{fig:ob_FS}(a,c), it is evident that the experimentally observed nodeless gap at $\beta$ pocket is a joint effort of the interlayer pairing within both $d_{x^2-y^2}$ and $d_{z^2}$ orbitals.
Finally, concerning the inplane inter-orbital pairing $\Delta_{||}^{xz}$, surprisedly, we find it actually forms a $d$-wave and can even enhance the dominated $\Delta_{\bot}^z$.
This is a unique feature for LNO tetrahedral lattice where the two distinct pairings can coexist within $A_{1g}$ channel, see  Appendix and Ref.~\cite{feng2025} for details. 

\section{\label{sec:5}Temperature dependence of superconducting gap}
\begin{figure}[t]
	\centering
	\includegraphics[width=1.\linewidth]{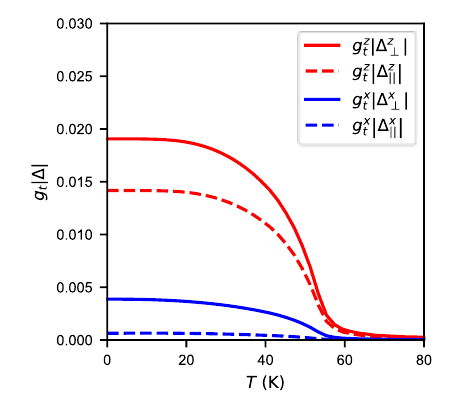}
        \caption{Superconducting gap as a function of temperature for different pairing bonds. $g_t^{z/x}$ is the renormalization factor for $d_{z^2}/d_{x^2-y^2}$ orbital, and $\Delta_{\bot/||}^{z/x}$ is the interlayer/inplane pairing bond for $d_{z^2}/d_{x^2-y^2}$ orbital.}
	\label{fig:Tc}
\end{figure}

% {\it Transition temperature.}$-$
In Fig.~\ref{fig:Tc}, we show the superconducting gap as a function of temperature $T$, which reveals two dominated pairing bonds from $d_{z^2}$ orbitals (red), in line with Fig.~\ref{fig:ob_FS}.
As increasing $T$, these gaps decrease in a mean-field manner and simultaneously drop to zero at $T_c$$\approx$60\ K.
Such value is comparable to the experimental reported $T_c$=40$\sim$60\ K in the thin films~\cite{ko_signatures_2025, zhou_ambient-pressure_2025, Osada_2025}.
Given that our previous RMFT study also obtains an experimental comparable $T_c\approx80$\ K  for the pressurized bulk LNO~\cite{luo2024high}, it is reasonable to believe that the decrease of $T_c$ in the thin films should originate from the decrease of $J_\bot$. 
This speculation draws us to the difference in the electronic structure, especially in the mainly concerned interlayer hopping $t_\bot^z$ for $d_{z^2}$ orbital.
Notably, we find that the $J_\bot$=0.135\ eV in this study is $\sim25\%$ smaller than that of $J_\bot$=0.18\ eV for the pressurized bulk~\cite{wu2024superexchange} as estimated from ED, which is coordinated with a simple estimation using $J_\bot\propto (t_\bot^z)^2$~\cite{luo2023bilayer,hu_electronic_2025}.
Recently, a strain-tuning experiment in LNO thin films reports an increase of $T_c$ from 10\ K in the tensile-strained $\mathrm{SrTiO_3}$ substrate to 60\ K in the compressively strained $\mathrm{LaAlO_3}$ substrate under 20\ GPa of pressure~\cite{Osada_2025}.
Recently, other groups also reported $T_c$ over 60 K in bilayer nickelate thin films through applying hydrostatic pressure~\cite{Lienhanced2025,Zhaopressure2026}.
This indicates a positive dependence of $T_c$ on the lattice ratio $c/a$ and, at first glance, violates the above understanding. 
%long-established viewpoint that $J_\bot$ should be responsible for the superconductivity~\cite{oh2023type, lu2024interlayer, qu2024bilayer}.
However, we tend to believe $J_\bot$ can only determine the upper bound of $T_c$, while other factors like hybridization, density wave, and interlayer Josephson coupling can also have intricate impact, as also pointed out in Ref.~\cite{qin_high-t_c_2023, chen_evidence_2024, meng_density-wave-like_2024, zhang2025pairingmechanismsuperconductivitypressurized, khasanov_pressure-enhanced_2025, Osada_2025, zheng2025}.

%We note that the calculated $T_c$ from our RMFT could be very sensitive to the magnitude of superexchanges, 

%There are two main branches, which are interlayer and intralayer $d_{z^2}$ orbitals $s\pm$ pairing, while $d$ wave pairing is totally surppress consitent with previous works~\cite{luo2024high,liu2023s}. We can see that the superconducting energy gap mainly contribute to $d_{z^2}$ orbitals, especially interlayer $d_{z^2}$ orbitals which come from $J^z_{\bot}$. 
%As temperature $T$ increases, all $g_t|\Delta|$ begin to decrease rapidly as $T$ reach about 40 K. The calculated critical temperature $T_c$ is comparable to experiments~\cite{ko_signatures_2025,zhou_ambient-pressure_2025}. 
%Given all these, we believe that our model and method can grasp the physics in LNO thin-film superconductivity.
%{\color{red} The $T_c$ result of RMFT is sensitive to superexchange coupling and electron number, which show that RMFT may not be a quantitatively correct way to calculate $T_c$, for RMFT is originally used for zero temperature~\cite{zhang_renormalised_1988}. }

\section{\label{sec:6}Discussion and conclusion}
% {\it Discussion.}$-$
%{{\color{red} We discuss more about our RMFT method for a better understanding of current result.
%The early RMFT is formulated for a single band t-J model in cuprate~\cite{zhang_renormalised_1988} in the frame of Gutzwiller projection. 
%As a strong coupling approach, the renormalization factors are estimated from the Gutzwiller approximation, which, in essense, is to de

%i.e., $\langle c_i^\dagger c_j\rangle_G/\langle c_i^\dagger c_j \rangle_0 =g_t^{ij}$, which, in essense, expresses  $\langle c_i^\dagger c_j\rangle= g_t\langle c_i^\dagger c_j \rangle_0$
%}

Our RMFT study of the superconductivity in LNO thin films has predicted an $s_\pm$-wave pairing symmetry.
The result largely resembles that of the pressurized bulk LNO~\cite{yang2023possible,liu2023s,zhang2024structural,luo2024high} giving the similarity in the electronic structure.
Mathematically, our application of Gutzwiller projection ~\cite{zhang_renormalised_1988,yang2023possible} on a multi-orbital $t-J$ model imposes a constraint on the renormalization condition 
to make the renormalization factor depends only on the orbital densities~\cite{luo2024high,edegger_gutzwillerrvb_2007}, so the RMFT formation can be solved self-consistently like the standard Bardeen-Cooper-Schrieffer (BCS) theory.
% {\color{red} to make the renormalization factor depends only on the orbital densities~\cite{luo2024high,edegger_gutzwillerrvb_2007}, so the RMFT formation can be solved self-consistently like the standard Bardeen-Cooper-Schrieffer (BCS) theory.} 
Therefore, it is expected that the obtained $s_\pm$-wave mainly reflects pairing preference for the LNO FS topology in the presence of $J_\bot$~\cite{Hujp2012}. 
Nevertheless, we find that the $s_\pm$-wave is quite robust under various approaches from weak to strong couplings, which convinces us that the fluctuation of correlated atomic states might not significantly alter such scenario.
In this scenario, the $\gamma$ pocket plays a central role.
Experimentally, there are two relevant ARPES measurements in which the $\gamma$ pocket shows visible profile for that performed in the normal state~\cite{li2025photoemissionevidencemultiorbitalholedoping} but invisible for that in the superconducting state~\cite{wang2025electronicstructurecompressivelystrained}.
From our result, this contradiction might be due to the opening of a large gap that exceeds the detected energy range, as shown Fig.~\ref{fig:gap}(d). More experimental evidences are expected on this issue.
Regarding the role of $J_{H}$, no significant impact is observed from our approach at the singlet pairing channel.
% {\color{red} at the singlet pairing channel}. 
We see that $J_H$ mainly contributes to an effective interlayer correlation of $d_{x^2-y^2}$ orbitals.
% {\color{red}We see that $J_H$ mainly contributes to an effective interlayer correlation of $d_{x^2-y^2}$ orbitals.}

However, in real materials, many factors will influence the pairing symmetry in La$_3$Ni$_2$O$_7$ thin films, such as  substrate-induced strain, oxygen vacancies and interfacial effects.
These factors are expected to influence the pairing symmetry through modifications to the electronic structure, such as the Fermi surface, hopping amplitudes, and superexchange couplings.
Substrate-induced compressive strain suppresses the $\gamma$ pocket and increases the c/a ratio~\cite{Zhaoelectronic2025}, which reduces both the interlayer $d_{z^2}$ hopping and corresponding superexchange couplings~\cite{hu_electronic_2025}.
Meanwhile, apical oxygen vacancies disrupt the Ni–O–Ni bonds along the $c$-axis, weakening the interlayer $d_{z^2}$ superexchange coupling.
Conversely, Sr diffusion from the substrate effectively introduce hole doping, which promotes the appearance of the $\gamma$ pocket~\cite{Liangle2025,Litheoretical2025}.
Within the RMFT framework, the absence of the $\gamma$ pocket drives the $d_{z^2}$ orbital toward half-filling, suppressing the interlayer $s_{\pm}$ gap, and the reduction in $J_{\bot}$ further diminishes the interlayer $s_\pm$ pairing tendency.

To summarize, we perform a RMFT study of the superconductivity in the LNO thin films using a bilayer $t-J$ model. Our result reveals an $s_\pm$-wave pairing symmetry driven by the interlayer superexchange coupling, resembling the bulk LNO case. 
Our result also roughly reproduces the experimentally reported nodeless shape of the superconducting gap at $\beta$ pocket ~\cite{shen2025anomalousenergygapsuperconducting}  as well as the superconducting $T_c$.
In addition, we systematically analyse the relation between orbital-resolved pairing bonds and the gap structure on FS. The result shows that the nodeless gap of $\beta$ pocket is related to the interlayer pairing within both $d_{z^2}$ and $d_{x^2-y^2}$ orbitals.
We also find the formation of inplane inter-orbital $d$-wave pairing between $d_{x^2-y^2}$ and $d_{z^2}$ orbitals that can enhance the dominated interlayer $s_\pm$-wave.
The coexistence is robust symmetrically in $\mathrm{La_3Ni_2O_7}$ lattice.
%This highlights the importance of $d_{x^2-y^2}$ orbital in involving the superconducting coherence throguth hybridization and the Hund's coupling.
%Our investigation shets light on the superconducting mechanism in LNO sytem for future exploration on the LNO system.

\begin{comment}
In this work, we construct a bilayer $t-J$ model to explore the superconducting properties in LNO thin film. 
Using exact diagnalization, we estimate the value of interlayer $d_{z^2}$ orbital spin interaction, which is 0.135 eV.
Through renomalized mean-field theory,  we calculate the pairing order parameters with different pairing symmetry at various temperatures. With slightly electron doping, the results appear to support $s\pm$ pairing symmetry and give a $T_c$ over 40 K. We show the  momentum distribution of energy gap on FS, where the energy gap on $\beta$ pocket is comparable to the experiment. Our results provide a method to understand superconductivity in LNO thin films in terms of $t-J$ model.
\end{comment}

We thank the useful discussion with W\'ei W\'u, Guan-Hao Feng.
This project was supported by NSFC-12494591, NSFC-92165204, NKRDPC-2022YFA1402802, Guangdong Provincial Key Laboratory of Magnetoelectric Physics and Devices (2022B1212010008), Guangdong Fundamental Research Center for Magnetoelectric Physics (2024B0303390001), and Guangdong Provincial Quantum Science Strategic Initiative (GDZX2401010).

 %Work at Sun Yat-Sen University was supported by the National Key Research and Development Program of China (Grants No. 2022YFA1402802, 2018YFA0306001), the National Natural Science Foundation of China (Grants No. 92165204, No.12174454, No. 11974432, No.12274472),  the Guangdong Basic and Applied Basic Research Foundation (Grants No. 2022A1515011618, No. 2021B1515120015), Guangdong Provincial Key Laboratory of Magnetoelectric Physics and Devices (Grant No. 2022B1212010008), Shenzhen International Quantum Academy (Grant No. SIQA202102), and Leading Talent Program of Guangdong Special Projects (201626003). 

\begin{appendix}
\section{Details on the pairing configurations for the bilayer $t-J$ model}

\begin{figure}
    \centering
    \includegraphics[width=1\linewidth]{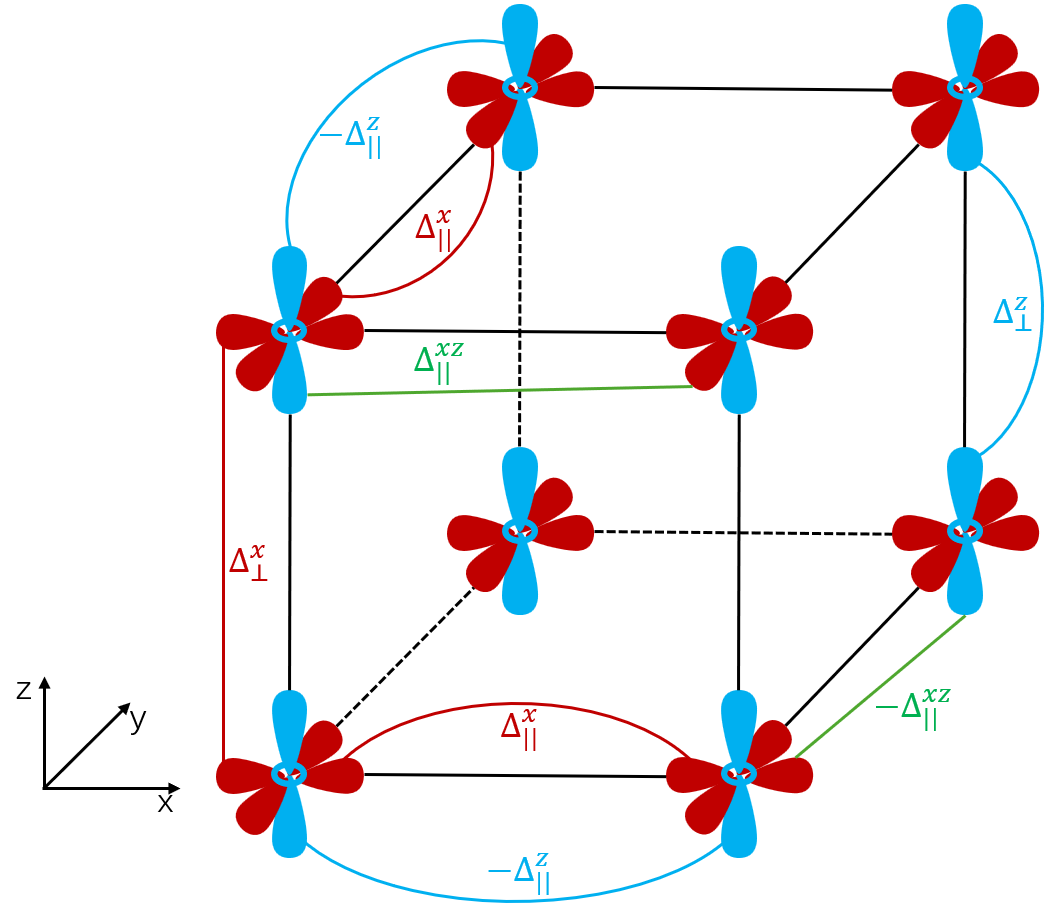}
    \caption{Pairing configuration of $\mathrm{La_3Ni_2O_7}$ thin films.}
    \label{fig:smFig1}
\end{figure}

\begin{figure}
    \centering
    \includegraphics[width=0.7\linewidth]{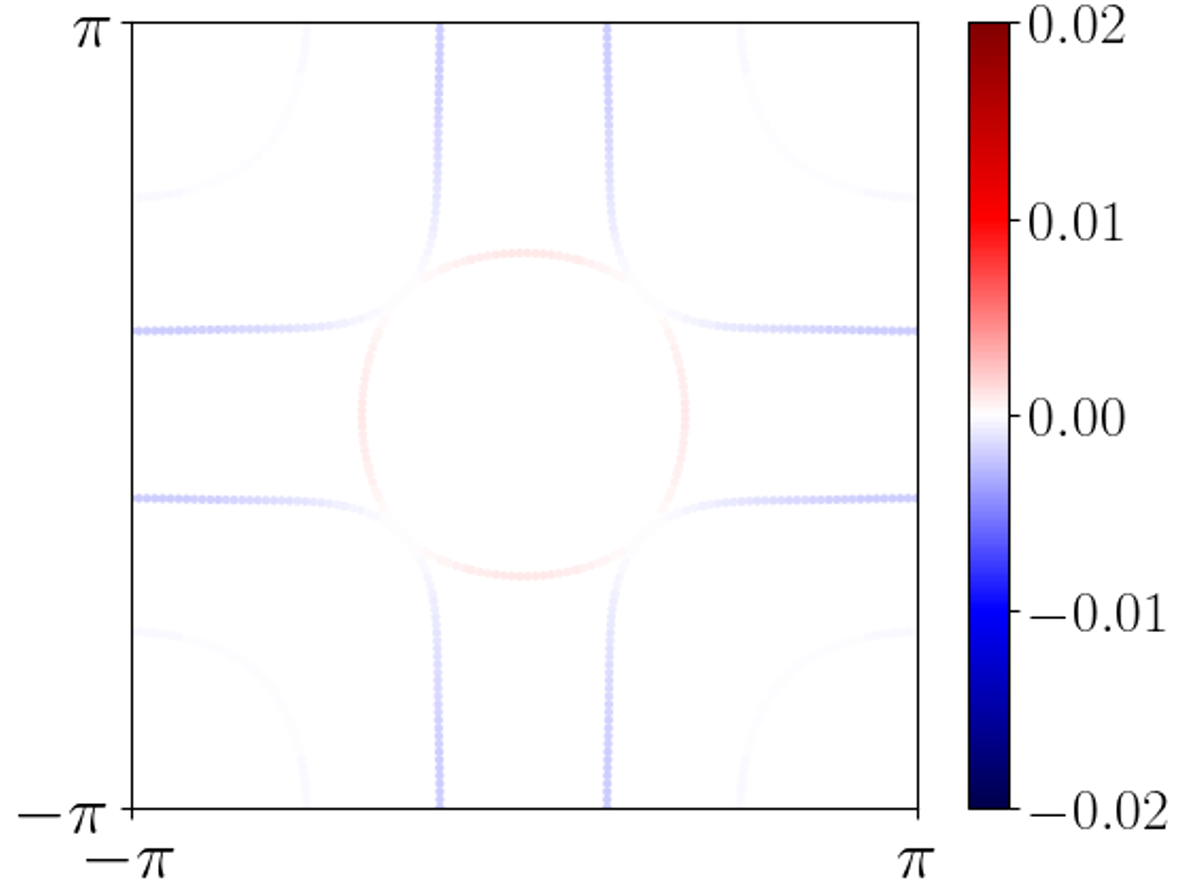}
    \caption{Superconducting gap $g_t^{xz}\Delta_{||}^{xz}$ projected onto FS.}
    \label{fig:smFig2}
\end{figure}

In this section, we illustrate the role of different superconducting pairing bonds in our bilayer $t-J$ model in $\mathrm{La_3Ni_2O_7}$ thin films.
The bilayer two-orbital model in our main text is defined under the basis 
$\Psi_{{\rm k}}=(d_{z_{1}},d_{x_{1}},d_{z_{2}},d_{x_{2}})^{T}$, and the extension to the Numbu basis becomes
\begin{align}
\mathcal{H}^{sc} & =\sum_{{\rm k}}\Phi_{{\rm k}}^{\dagger}H_{{\rm k}}^{sc}\Phi_{{\rm k}},\\
H_{{\rm k}}^{sc} & =\left(\begin{array}{cccc}
H_{||} & H_{\bot} & \Delta_{||} & \Delta_{\bot}\\
H_{\bot} & H_{||} & \Delta_{\bot} & \Delta_{||}\\
\Delta_{||} & \Delta_{\bot} & -H_{||} & -H_{\bot}\\
\Delta_{\bot} & \Delta_{||} & -H_{\bot} & -H_{||}
\end{array}\right)_{{\rm }}.
\end{align}
Here $\Phi=(\Psi_{{\rm }}^{T},\Psi^{\dagger})^{T}=\left(d_{z_{1}},d_{x_{1}},d_{z_{2}},d_{x_{2}},d_{z_{1}}^{\dagger},d_{x_{1}}^{\dagger},d_{z_{2}}^{\dag},d_{x_{2}}^{\dag}\right)^{T}$,
and $H_{||/\bot}$, $\Delta_{||/\bot}$ are the $2\times2$ matrices
with the subscript denoting the intralayer/interlayer. 
%Using the notation in Ref.~\cite{luo2023bilayer}, 
$H_{||},H_\bot$ are obtained by taking the averge over the double stacks for the one-UC model of $\mathrm{La_3Ni_2O_7}$ thin films in Ref.~\cite{hu_electronic_2025}, which is expressed as
\begin{align}
H_{||}=\left(\begin{array}{cc}
T_{{\rm {\rm k}}}^{z} & V_{{\rm k}}\\
V_{{\rm k}} & T_{{\rm {\rm k}}}^{x}
\end{array}\right),\ H_{\bot}=\left(\begin{array}{cc}
T_{\bot,{\rm {\rm k}}}^{z} & V_{{\rm k}}^{\prime}\\
V_{{\rm k}}^{\prime} & T_{\bot,{\rm {\rm k}}}^{x}
\end{array}\right),
\end{align}
with
\begin{widetext}
\begin{align}
    T_{{\rm {\rm k}}}^{z} &= 
    -0.217\left(\cos  {k}_{x}+\cos {k}_{y}\right) 
    -0.073\cos {k}_{x}\cos {k}_{y} 
    -0.021\left(\cos  2{k}_{x}+\cos 2{k}_{y}\right) 
    -0.005\left(\cos  3{k}_{x}+\cos 3{k}_{y}\right) + 0.431, \nonumber\\
    T_{{\rm {\rm k}}}^{x} &= 
    -0.922\left(\cos  {k}_{x}+\cos {k}_{y}\right) +
    0.301\cos {k}_{x}\cos {k}_{y} -0.108\left(\cos  2{k}_{x}+\cos 2{k}_{y}\right) 
    -0.025\left(\cos  3{k}_{x}+\cos 3{k}_{y}\right) 
    + 0.881, \nonumber\\
    T_{\bot,{\rm {\rm k}}}^{z} &=
    -0.550 + 0.041\left(\cos {k}_{x}+\cos {k}_{y}\right), \\
    T_{\bot,{\rm {\rm k}}}^{x} &=
    0.005,\nonumber \\
    V_{{\rm k}} &= 0.429\left(\cos {k}_{x}-\cos {k}_{y}\right) + 
    0.041\left(\cos 2{k}_{x}-\cos 2{k}_{y}\right),\nonumber\\
    V_{{\rm k}}^{\prime} &=
    -0.061\left(\cos {k}_{x}-\cos {k}_{y}\right).\nonumber
\end{align}
\end{widetext}
Meanwhile, our RMFT solution of $\mathrm{La_3Ni_2O_7}$ thin films shows that
\begin{align}
\label{eq4}
\Delta_{||}&=\Delta_{||}^{A_{1g}}+\Delta^{B_{1g}}_{||},\\
\Delta_{||}^{A_{1g}}&= 2\left(\begin{array}{cc}
-\Delta_{||}^{z}(\cos k_{x}+\cos k_{y}) & \Delta_{||}^{xz}(\cos k_{x}-\cos k_{y})\\
\label{eq5}
\Delta_{||}^{xz}(\cos k_{x}-\cos k_{y}) & \Delta_{||}^{x}(\cos k_{x}+\cos k_{y})
\end{array}\right)_{{\rm }},\\
\Delta_{\bot}&=\left(\begin{array}{cc}
\Delta_{\bot}^{z} & 0\\
0 & \Delta_{\bot}^{x}
\end{array}\right)_{{\rm }}.
\end{align}
Here, $\Delta_{||}^z$,$\Delta_{||}^x$,$\Delta_{||}^{xz}$, $\Delta_{\bot}^{z}$,$\Delta_{\bot}^{x}$ are all positive.
The pairing configuration is shown in Fig.~\ref{fig:smFig1}.
In Eq.~(\ref{eq4}) we use $A_{1g}$, $B_{1g}$  notations to denote the pairing symmetry instead of the commonly used $s_\pm,d$ symbols, which are more general and can aviod ambiguity in a multi-orbital system.
The presence of $\Delta_{||}^{B_{1g}}$ is due to the consideration of $J_{||},J_{xz}$ in our RMFT, which is  marginal and can be safely discarded from the result.
For the expression of $\Delta_{||}^{A_{1g}}$ in Eq.~(\ref{eq5}), it is quite unexpected that, besides the diagonal terms that hold the $s_\pm$-wave form factor, the $d$-wave form factor appears in the off-diagonal term relating the inplane pairing between $d_{z^2}$ and $d_{x^2-y^2}$ orbitals.
The coexisting $s_\pm,d$ waves within $A_{1g}$ channel
should be better understood in terms of group theory, and we suggest reader to Ref.~\cite{feng2025} where a full topological classification of the superconducting pairings in $\mathrm{La_3Ni_2O_7}$ system is made.
One might expect a competitive relation of the two pairing symmetries.
To better illustrate, in Fig.~\ref{fig:smFig2} we show the projection of $g_t^{xz}\Delta_{||}^{xz}$ onto Fermi surface. Surprisedly, it demonstrates a phase structure matching that in Fig. 3(a) of the main text, i.e., it  amplifies the gaps at $\alpha,\beta$ pockets and can enhance the dominated $\Delta_{\bot}^z$.
But for the diagonal $\Delta_{||}^z,\Delta_{||}^x$, as discussed in the main text, cancellation of the gap in certain region of $\alpha,\beta$ pockets occurs with respect to $\Delta_{\bot}^z$, which inherently reflects the ``pairing frustration". We note that the above method in determining the relation of different pairing bonds is equivalent to minimizing the free energy functional $F[\Delta_1,\Delta_2,...]=-k_{\rm B}T\ln {\rm Tr}e^{-H/k_{\rm B}T}$, which encodes all possible relations of different pairings $\Delta_1,\Delta_2,...$  as frustration, competition and cooperation.
In principle, one can always constraint the calculation of $F$ to certain pairing symmetry $\Delta_i$ and make a comparison of the energy difference. This standard approach is apparently very tedious and less intuitive.
In our self-consistent approach, with no constraint on the symmetry, corresponding to a full optimization of $F$ with respect to all relevant $\Delta_i$, we find the relation of different pairing channels naturally manifests as the sign structure of the gaps.
In case of $\mathrm{La_3Ni_2O_7}$, it is due to the large $J_\bot$ that $F$ will priority gain tremendous energy by forming $\Delta^z_\bot$, after that other $\Delta_i$ will form on top of that to see if it can further decrease $F$. This naturally explains the outcome of Fig.~2 in the main text.

To proceed with, we perform a transform to the bonding-antibonding representation as
\begin{align}
H_{\pm}=H_{||}\pm H_{\bot},\ \Delta_{\pm}=\Delta_{||}\pm\Delta_{\bot}.
\end{align}
It is clear from this representation that, when along the Brillouin
zone diagonal direction, the gaps for $\alpha,\beta,\gamma$ pockets
respectively acquire the form 
\begin{align}
\label{eq11}
2\Delta_{||}^{x}(\cos k_{x}+\cos k_{y})+\Delta_{\bot}^{x}, \\
\label{eq12}
2\Delta_{||}^{x}(\cos k_{x}+\cos k_{y})-\Delta_{\bot}^{x}, \\
\label{eq13}
-2\Delta_{||}^{z}(\cos k_{x}+\cos k_{y})+\Delta_{\bot}^{z}.
\end{align}

Eq.~(\ref{eq13}) shows that, in order to maximize the gap at $\gamma$
pocket around M=($\pi$,$\pi$) point, $-\Delta_{||}^{z}$, $\Delta_{\bot}^{z}$
have to be in antiphase.
Such antiphase is also identified in Ref.~\cite{lu2024interlayer}, but there only $d_{x^2-y^2}$ orbital is relevant while $d_{z^2}$ orbital has been integrated out a priori.
But for $d_{x^2-y^2}$ orbital in our case, as demonstrated in Eq.~(\ref{eq11}), they actually hold the same sign in order to maximize the gap at $\alpha$ pocket.
For $\beta$ pocket in Eq.~(\ref{eq12}), a competition between $\Delta_{||}^x,\Delta_{\bot}^x$ occurs, and
our result as well as the experimental evidence suggest that $4\Delta_{||}^x\cos k_x<\Delta_{\bot}^x$.
We also note that in Ref.~\cite{yang2023possible}, a negative sign of $\Delta_{||}^x$ is revealed using the functional renormalization group calculation for a multiorbital atomic Coulomb interaction. Since the pairing configuration in Fig.~\ref{fig:smFig1} mainly reflects the pairing preference for the $\mathrm{La_3Ni_2O_7}$ FS in the presence of  $J_\bot$, the slight discrepancy may be attributed to other high-order processes besides $J_\bot$.

\end{appendix}

\bibliography{ref}

@misc{hu_electronic_2025,
    title={Electronic structures and multi-orbital models of {La$_3$Ni$_2$O$_7$} thin films at ambient pressure}, 
    journal={Commun. Phys.},
    volume={8},
    pages = {506},
   doi={https://doi.org/10.1038/s42005-025-02411-8},
   number={1}, 
   publisher={Springer Science and Business Media LLC},
   author={Hu, Xunwu and Qiu, Wenyuan and Chen, Cui-Qun and Luo, Zhihui and Yao, Dao-Xin},
   year={2025}
}

@misc{feng2025,
      author={Guan-Hao Feng and Jun Quan and Yusheng Hou},
      year={2025},
      title={Unconventional Superconductivity in $\mathrm{La_{3}Ni_{2}O_{7}}$ from the Perspective of Symmetry}, 
      eprint={2506.01764},
      archivePrefix={arXiv},
}

@article{gutz,
    title = {Correlation of {Electrons} in a {Narrow} $s$ {Band}},
    volume = {137},
    url = {https://link.aps.org/doi/10.1103/PhysRev.137.A1726},
    doi = {10.1103/PhysRev.137.A1726},
    number = {6A},
    urldate = {2025-06-25},
    journal = {Physical Review},
    author = {Gutzwiller, Martin C.},
    month = mar,
    year = {1965},
    pages = {A1726--A1735},
}

@article{shen2025anomalousenergygapsuperconducting,
      author={Jianchang Shen and Yu Miao and Zhipeng Ou and Guangdi Zhou and Yaqi Chen and Runqing Luan and Hongxu Sun and Zikun Feng and Xinru Yong and Peng Li and Yueying Li and Lizhi Xu and Wei Lv and Zihao Nie and Heng Wang and Haoliang Huang and Yu-Jie Sun and Qi-Kun Xue and Zhuoyu Chen and Junfeng He},
      title={Nodeless superconducting gap and electron-boson coupling in {(La,Pr,Sm)$_{3}$Ni$_2$O$_7$} films}, 
    number={6805},
   journal={Science},
   year={2026},
   month={June}, 
   pages={1396–1400},
    DOI={10.1126/science.adw8329},
    volume = {392}
}

@misc{fan2025superconductinggapsrevealedstm,
      author={Shengtai Fan and Mengjun Ou and Marius Scholten and Qing Li and Zhiyuan Shang and Yi Wang and Jiasen Xu and Huan Yang and Ilya M. Eremin and Hai-Hu Wen},
      year={2025},
      title={Superconducting gap structure and bosonic mode in La$_2$PrNi$_2$O$_7$ thin films at ambient pressure},
      eprint={2506.01788},
      archivePrefix={arXiv},
}

@article{Hujp2012,
   author = {Hu, Jiangping and Ding, Hong},
   title = {Local antiferromagnetic exchange and collaborative Fermi surface as key ingredients of high temperature superconductors},
   journal = {Sci. Rep.},
   volume = {2},
   number = {1},
   ISSN = {2045-2322},
   url = {https://doi.org/10.1038/srep00381},
   year = {2012},
   type = {Journal Article}
}

@article{zhang_renormalised_1988,
	title = {A renormalised {Hamiltonian} approach to a resonant valence bond wavefunction},
	volume = {1},
	issn = {0953-2048, 1361-6668},
	url = {https://iopscience.iop.org/article/10.1088/0953-2048/1/1/009},
	doi = {10.1088/0953-2048/1/1/009},
	number = {1},
	journal = {Superconductor Science and Technology},
	author = {Zhang, F C and Gros, C and Rice, T M and Shiba, H},
	month = jun,
	year = {1988},
	pages = {36--46},
}

@article{li2025ambientpressuregrowthbilayer,
      title={Bulk superconductivity up to 96 {K} in pressurized nickelate single crystals},
   volume={649},
   DOI={10.1038/s41586-025-09954-4},
   number={8098},
   journal={Nature},
   author={Li, Feiyu and Xing, Zhenfang and Peng, Di and Dou, Jie and Guo, Ning and Ma, Liang and Zhang, Yulin and Wang, Lingzhen and Luo, Jun and Yang, Jie and Zhang, Jian and Chang, Tieyan and Chen, Yu-Sheng and Cai, Weizhao and Cheng, Jinguang and Wang, Yuzhu and Liu, Yuxin and Luo, Tao and Hirao, Naohisa and Matsuoka, Takahiro and Kadobayashi, Hirokazu and Zeng, Zhidan and Zheng, Qiang and Zhou, Rui and Zeng, Qiaoshi and Tao, Xutang and Zhang, Junjie},
   year={2025},
   month={Dec}, 
   pages={871–878} 
}

@article{Osada_2025,
   title={Strain-tuning for superconductivity in {La}$_3${Ni}$_2${O}$_7$ thin films},
   volume={8},
   ISSN={2399-3650},
   url={http://dx.doi.org/10.1038/s42005-025-02154-6},
   number={1},
   journal={Commun. Phys.},
   author={Osada, Motoki and Terakura, Chieko and Kikkawa, Akiko and Nakajima, Masamichi and Chen, Hsiao-Yi and Nomura, Yusuke and Tokura, Yoshinori and Tsukazaki, Atsushi},
   year={2025},
   month=jun 
}

@article{zheng2025,
	title = {$s\pm$-wave superconductivity in the bilayer two-orbital {Hubbard} model},
	volume = {111},
	url = {https://link.aps.org/doi/10.1103/PhysRevB.111.035108},
	doi = {10.1103/PhysRevB.111.035108},
	number = {3},
	urldate = {2025-06-25},
	journal = {Phys. Rev. B},
	author = {Zheng, Yao-Yuan and Wú, Wéi},
	month = jan,
	year = {2025},
	pages = {035108},
}

@article{qin_high-t_c_2023,
	title = {High-${T}_{c}$ superconductivity by mobilizing local spin singlets and possible route to higher ${T}_{c}$ in pressurized {La}$_3${Ni}$_2${O}$_7$},
	volume = {108},
	url = {https://link.aps.org/doi/10.1103/PhysRevB.108.L140504},
	doi = {10.1103/PhysRevB.108.L140504},
	number = {14},
	urldate = {2025-06-25},
	journal = {Phys. Rev. B},
	author = {Qin, Qiong and Yang, Yi-feng},
	month = oct,
	year = {2023},
	pages = {L140504},
}

@article{chen_evidence_2024,
	title = {Evidence of {Spin} {Density} {Waves} in {La}$_3${Ni}$_2${O}$_{7-\delta}$},
	volume = {132},
	url = {https://link.aps.org/doi/10.1103/PhysRevLett.132.256503},
	doi = {10.1103/PhysRevLett.132.256503},
	number = {25},
	urldate = {2025-06-18},
	journal = {Phys. Rev. Lett.},
	author = {Chen, Kaiwen and Liu, Xiangqi and Jiao, Jiachen and Zou, Muyuan and Jiang, Chengyu and Li, Xin and Luo, Yixuan and Wu, Qiong and Zhang, Ningyuan and Guo, Yanfeng and Shu, Lei},
	month = jun,
	year = {2024},
	pages = {256503},
}

@article{meng_density-wave-like_2024,
	title = {Density-wave-like gap evolution in {La}$_3${Ni}$_2${O}$_7$ under high pressure revealed by ultrafast optical spectroscopy},
	volume = {15},
	issn = {2041-1723},
	url = {https://www.nature.com/articles/s41467-024-54518-1},
	doi = {10.1038/s41467-024-54518-1},
	number = {1},
	urldate = {2025-06-18},
	journal = {Nat. Commun.},
	author = {Meng, Yanghao and Yang, Yi and Sun, Hualei and Zhang, Sasa and Luo, Jianlin and Chen, Liucheng and Ma, Xiaoli and Wang, Meng and Hong, Fang and Wang, Xinbo and Yu, Xiaohui},
	month = {nov},
	year = {2024},
	pages = {10408},
}

@article{khasanov_pressure-enhanced_2025,
	title = {Pressure-enhanced splitting of density wave transitions in {La}$_3${Ni}$_2${O}$_{7-\delta}$},
	volume = {21},
	issn = {1745-2481},
	url = {https://www.nature.com/articles/s41567-024-02754-z},
	doi = {10.1038/s41567-024-02754-z},
	number = {3},
	urldate = {2025-06-18},
	journal = {Nat. Phys.},
	author = {Khasanov, Rustem and Hicken, Thomas J. and Gawryluk, Dariusz J. and Sazgari, Vahid and Plokhikh, Igor and Sorel, Loïc Pierre and Bartkowiak, Marek and Bötzel, Steffen and Lechermann, Frank and Eremin, Ilya M. and Luetkens, Hubertus and Guguchia, Zurab},
	month = mar,
	year = {2025},
	pages = {430--436},
}

@misc{wang2025electronicstructurecompressivelystrained,
      author={Bai Yang Wang and Yong Zhong and Sebastien Abadi and Yidi Liu and Yijun Yu and Xiaoliang Zhang and Yi-Ming Wu and Ruohan Wang and Jiarui Li and Yaoju Tarn and Eun Kyo Ko and Vivek Thampy and Makoto Hashimoto and Donghui Lu and Young S. Lee and Thomas P. Devereaux and Chunjing Jia and Harold Y. Hwang and Zhi-Xun Shen},
      title={Electronic structure of compressively strained thin film {La$_2$PrNi$_2$O$_7$}},
      year={2025},
      eprint={2504.16372},
      archivePrefix={arXiv}
}

@article{pan_effect_2024,
	title = {Effect of {Rare}-{Earth} {Element} {Substitution} in {Superconducting} {R}$_3${Ni}$_2${O}$_{7}$ under {Pressure}},
	volume = {41},
	issn = {0256-307X},
	url = {https://dx.doi.org/10.1088/0256-307X/41/8/087401},
	doi = {10.1088/0256-307X/41/8/087401},
	number = {8},
	urldate = {2025-06-25},
	journal = {Chin. Phys. Lett.},
	author = {Pan, Zhiming and Lu, Chen and Yang, Fan and Wu, Congjun},
	month = aug,
	year = {2024},
	pages = {087401},
}

@article{zhang_trends_2023,
	title = {Trends in electronic structures and $s_{\pm}$-wave pairing for the rare-earth series in bilayer nickelate superconductor {R}$_3${Ni}$_2${O}$_{7}$},
	volume = {108},
	url = {https://link.aps.org/doi/10.1103/PhysRevB.108.165141},
	doi = {10.1103/PhysRevB.108.165141},
	number = {16},
	urldate = {2025-06-25},
	journal = {Phys. Rev. B},
	author = {Zhang, Yang and Lin, Ling-Fang and Moreo, Adriana and Maier, Thomas A. and Dagotto, Elbio},
	month = oct,
	year = {2023},
	pages = {165141},
}

@misc{zhang2025pairingmechanismsuperconductivitypressurized,
      author={Ming Zhang and Cui-Qun Chen and Dao-Xin Yao and Fan Yang},
      title={Pairing mechanism and superconductivity in pressurized {La}$_5${Ni}$_3${O}$_{11}$}, 
      year={2025},
      eprint={2505.15906},
      archivePrefix={arXiv},
}

@misc{shi2025effectcarrierdopingthickness,
      author={Haoliang Shi and Zihao Huo and Guanlin Li and Hao Ma and Tian Cui and Dao-Xin Yao and Defang Duan},
      year={2025},
      title={The effect of Carrier Doping and Thickness on the Electronic Structures of {La$_3$Ni$_2$O$_7$} Thin Films}, 
      eprint={2502.04255},
      archivePrefix={arXiv},
}

@article{edegger_gutzwillerrvb_2007,
	title = {Gutzwiller–{RVB} theory of high-temperature superconductivity: {Results} from renormalized mean-field theory and variational {Monte} {Carlo} calculations},
	volume = {56},
	issn = {0001-8732 1460-6976},
	doi = {10.1080/00018730701627707},
	number = {6},
	journal = {Advances in Physics},
	author = {Edegger, B. and Muthukumar, V. N. and Gros, C.},
	year = {2007},
	pages = {927--1033},
}

@article{Sunsignatures2023,
  title = {Signatures of superconductivity near 80 K in a nickelate under high pressure},
  author = {Sun, Hualei and Huo, Mengwu and Hu, Xunwu and Li, Jingyuan and Liu, Zengjia and Han, Yifeng and Tang, Lingyun and Mao, Zhongquan and Yang, Pengtao and Wang, Bosen and Cheng, Jinguang and Yao, Dao-Xin and Zhang, Guang-Ming and Wang, Meng},
  year = {2023},
  month = sep,
  journal = {Nature},
  volume = {621},
  number = {7979},
  pages = {493--498},
  doi = {10.1038/s41586-023-06408-7},
}

@article{luo2023bilayer,
  title = {Bilayer Two-Orbital Model of $\mathrm{L}{\mathrm{a}}_{3}\mathrm{N}{\mathrm{i}}_{2}\mathrm{O}_{7}$ under Pressure},
  author = {Luo, Zhihui and Hu, Xunwu and Wang, Meng and Wú, Wéi and Yao, Dao-Xin},
  journal = {Phys. Rev. Lett.},
  volume = {131},
  number = {12},
  pages = {126001},
  year = {2023},
  doi = {10.1103/PhysRevLett.131.126001}
}

@article{zhang2023electronic,
  title = {Electronic structure, dimer physics, orbital-selective behavior, and magnetic tendencies in the bilayer nickelate superconductor $\mathrm{La}_{3}\mathrm{Ni}_{2}\mathrm{O}_{7}$ under pressure},
  author = {Zhang, Yang and Lin, Ling-Fang and Moreo, Adriana and Dagotto, Elbio},
  journal = {Phys. Rev. B},
  volume = {108},
  number = {18},
  pages = {L180510},
  year = {2023},
  doi = {10.1103/PhysRevB.108.L180510}
}

@article{lechermann2023electronic,
  title = {Electronic correlations and superconducting instability in $\mathrm{L}{\mathrm{a}}_{3}\mathrm{N}{\mathrm{i}}_{2}\mathrm{O}_{7}$ under high pressure},
  author = {Lechermann, Frank and Gondolf, Jannik and Bötzel, Steffen and Eremin, Ilya M.},
  journal = {Phys. Rev. B},
  volume = {108},
  number = {20},
  pages = {L201121},
  year = {2023},
  doi = {10.1103/PhysRevB.108.L201121}
}

@article{shilenko2023correlated,
  title = {Correlated electronic structure, orbital-selective behavior, and magnetic correlations in double-layer $\mathrm{La}_{3}\mathrm{Ni}_{2}\mathrm{O}_{7}$ under pressure},
  author = {Shilenko, D. A. and Leonov, I. V.},
  journal = {Phys. Rev. B},
  volume = {108},
  number = {12},
  pages = {125105},
  year = {2023},
  doi = {10.1103/PhysRevB.108.125105}
}

@article{chen2025charge,
  title = {Charge and spin instabilities in superconducting $\mathrm{La}_{3}\mathrm{Ni}_{2}\mathrm{O}_{7}$},
  author = {Chen, Xuejiao and Jiang, Peiheng and Li, Jie and Zhong, Zhicheng and Lu, Yi},
  journal = {Phys. Rev. B},
  volume = {111},
  number = {1},
  pages = {014515},
  year = {2025},
  doi = {10.1103/PhysRevB.111.014515}
}

@article{ouyang2024absence,
  title = {Absence of electron-phonon coupling superconductivity in the bilayer phase of $\mathrm{La}_{3}\mathrm{Ni}_{2}\mathrm{O}_{7}$ under pressure},
  author = {Ouyang, Zhenfeng and Gao, Miao and Lu, Zhong-Yi},
  journal = {npj Quantum Mater.},
  volume = {9},
  number = {1},
  pages = {1-6},
  year = {2024},
  doi = {10.1038/s41535-024-00689-5}
}

@article{liu2023s,
  title = {${s}^{\ifmmode\pm\else\textpm\fi{}}$-Wave Pairing and the Destructive Role of Apical-Oxygen Deficiencies in $\mathrm{La}_{3}\mathrm{Ni}_{2}\mathrm{O}_{7}$ under Pressure},
  author = {Liu, Yu-Bo and Mei, Jia-Wei and Ye, Fei and Chen, Wei-Qiang and Yang, Fan},
  journal = {Phys. Rev. Lett.},
  volume = {131},
  number = {23},
  pages = {236002},
  year = {2023},
  doi = {10.1103/PhysRevLett.131.236002}
}

@article{heier2024competing,
  title = {Competing ${d}_{xy}$ and ${s}_{\ifmmode\pm\else\textpm\fi{}}$ pairing symmetries in superconducting $\mathrm{La}_{3}\mathrm{Ni}_{2}\mathrm{O}_{7}$: $\mathrm{LDA}+\mathrm{FLEX}$ calculations},
  author = {Heier, Griffin and Park, Kyungwha and Savrasov, Sergey Y.},
  journal = {Phys. Rev. B},
  volume = {109},
  number = {10},
  pages = {104508},
  year = {2024},
  doi = {10.1103/PhysRevB.109.104508}
}

@article{zhang2024structural,
  title = {Structural phase transition, s±-wave pairing, and magnetic stripe order in bilayered superconductor $\mathrm{La}_{3}\mathrm{Ni}_{2}\mathrm{O}_{7}$ under pressure},
  author = {Zhang, Yang and Lin, Ling-Fang and Moreo, Adriana and Maier, Thomas A. and Dagotto, Elbio},
  journal = {Nat. Commun.},
  volume = {15},
  number = {1},
  pages = {2470},
  year = {2024},
  doi = {10.1038/s41467-024-46622-z}
}

@article{zhang2024electronic,
  title = {Electronic structure, magnetic correlations, and superconducting pairing in the reduced Ruddlesden-Popper bilayer $\mathrm{La}_{3}\mathrm{Ni}_{2}\mathrm{O}_{6}$ under pressure: Different role of ${d}_{3{z}^{2}\ensuremath{-}{r}^{2}}$ orbital compared with $\mathrm{La}_{3}\mathrm{Ni}_{2}\mathrm{O}_{7}$},
  author = {Zhang, Yang and Lin, Ling-Fang and Moreo, Adriana and Maier, Thomas A. and Dagotto, Elbio},
  journal = {Phys. Rev. B},
  volume = {109},
  number = {4},
  pages = {045151},
  year = {2024},
  doi = {10.1103/PhysRevB.109.045151}
}

@article{christiansson2023correlated,
  title = {Correlated Electronic Structure of $\mathrm{La}_{3}\mathrm{Ni}_{2}\mathrm{O}_{7}$ under Pressure},
  author = {Christiansson, Viktor and Petocchi, Francesco and Werner, Philipp},
  journal = {Phys. Rev. Lett.},
  volume = {131},
  number = {20},
  pages = {206501},
  year = {2023},
  doi = {10.1103/PhysRevLett.131.206501}
}

@article{wu2024superexchange,
  title = {Superexchange and charge transfer in the nickelate superconductor $\mathrm{La}_{3}\mathrm{Ni}_{2}\mathrm{O}_{7}$ under pressure},
  author = {Wú, Wéi and Luo, Zhihui and Yao, Dao-Xin and Wang, Meng},
  journal = {Sci. China Phys. Mech. Astron.},
  volume = {67},
  number = {11},
  pages = {117402},
  year = {2024},
  doi = {10.1007/s11433-023-2300-4}
}

@article{tian2024correlation,
  title = {Correlation effects and concomitant two-orbital ${s}_{\ifmmode\pm\else\textpm\fi{}}$-wave superconductivity in $\mathrm{La}_{3}\mathrm{Ni}_{2}\mathrm{O}_{7}$ under high pressure},
  author = {Tian, Yi-Heng and Chen, Yin and Wang, Jia-Ming and He, Rong-Qiang and Lu, Zhong-Yi},
  journal = {Phys. Rev. B},
  volume = {109},
  number = {16},
  pages = {165154},
  year = {2024},
  doi = {10.1103/PhysRevB.109.165154}
}

@article{ryee2024quenched,
  title = {Quenched Pair Breaking by Interlayer Correlations as a Key to Superconductivity in $\mathrm{La}_{3}\mathrm{Ni}_{2}\mathrm{O}_{7}$},
  author = {Ryee, Siheon and Witt, Niklas and Wehling, Tim O.},
  journal = {Phys. Rev. Lett.},
  volume = {133},
  number = {9},
  pages = {096002},
  year = {2024},
  doi = {10.1103/PhysRevLett.133.096002}
}

@article{zhang2024strong,
  title = {Strong Pairing Originated from an Emergent ${Z}_{2}$ Berry Phase in $\mathrm{La}_{3}\mathrm{Ni}_{2}\mathrm{O}_{7}$},
  author = {Zhang, Jia-Xin and Zhang, Hao-Kai and You, Yi-Zhuang and Weng, Zheng-Yu},
  journal = {Phys. Rev. Lett.},
  volume = {133},
  number = {12},
  pages = {126501},
  year = {2024},
  doi = {10.1103/PhysRevLett.133.126501}
}

@article{shen2023effective,
  title = {Effective Bi-Layer Model Hamiltonian and Density-Matrix Renormalization Group Study for the High-Tc Superconductivity in $\mathrm{La}_{3}\mathrm{Ni}_{2}\mathrm{O}_{7}$ under High Pressure},
  author = {Shen, Yang and Qin, Mingpu and Zhang, Guang-Ming},
  journal = {Chin. Phys. Lett.},
  volume = {40},
  number = {12},
  pages = {127401},
  year = {2023},
  doi = {10.1088/0256-307X/40/12/127401}
}

@article{lu2024interlayer,
  title = {Interlayer-Coupling-Driven High-Temperature Superconductivity in $\mathrm{La}_{3}\mathrm{Ni}_{2}\mathrm{O}_{7}$ under Pressure},
  author = {Lu, Chen and Pan, Zhiming and Yang, Fan and Wu, Congjun},
  journal = {Phys. Rev. Lett.},
  volume = {132},
  number = {14},
  pages = {146002},
  year = {2024},
  doi = {10.1103/PhysRevLett.132.146002}
}

@article{oh2023type,
  title = {Type-{II} $t-{J}$ model and shared superexchange coupling from Hund's rule in superconducting $\mathrm{La}_{3}\mathrm{Ni}_{2}\mathrm{O}_{7}$},
  author = {Oh, Hanbit and Zhang, Ya-Hui},
  journal = {Phys. Rev. B},
  volume = {108},
  number = {17},
  pages = {174511},
  year = {2023},
  doi = {10.1103/PhysRevB.108.174511}
}

@article{qu2024bilayer,
  title = {Bilayer ${t\text{\ensuremath{-}}J\text{\ensuremath{-}}J}_{\ensuremath{\perp}}$ Model and Magnetically Mediated Pairing in the Pressurized Nickelate $\mathrm{La}_{3}\mathrm{Ni}_{2}\mathrm{O}_{7}$},
  author = {Qu, Xing-Zhou and Qu, Dai-Wei and Chen, Jialin and Wu, Congjun and Yang, Fan and Li, Wei and Su, Gang},
  journal = {Phys. Rev. Lett.},
  volume = {132},
  number = {3},
  pages = {036502},
  year = {2024},
  doi = {10.1103/PhysRevLett.132.036502}
}

@article{yang2024strong,
  title = {Strong pairing from a small Fermi surface beyond weak coupling: Application to $\mathrm{La}_{3}\mathrm{Ni}_{2}\mathrm{O}_{7}$},
  author = {Yang, Hui and Oh, Hanbit and Zhang, Ya-Hui},
  journal = {Phys. Rev. B},
  volume = {110},
  number = {10},
  pages = {104517},
  year = {2024},
  doi = {10.1103/PhysRevB.110.104517}
}

@article{fan2024superconductivity,
  title = {Superconductivity in nickelate and cuprate superconductors with strong bilayer coupling},
  author = {Fan, Zhen and Zhang, Jian-Feng and Zhan, Bo and Lv, Dingshun and Jiang, Xing-Yu and Normand, Bruce and Xiang, Tao},
  journal = {Phys. Rev. B},
  volume = {110},
  number = {2},
  pages = {024514},
  year = {2024},
  doi = {10.1103/PhysRevB.110.024514}
}

@article{liao2023electron,
  title = {Electron correlations and superconductivity in $\mathrm{La}_{3}\mathrm{Ni}_{2}\mathrm{O}_{7}$ under pressure tuning},
  author = {Liao, Zhiguang and Chen, Lei and Duan, Guijing and Wang, Yiming and Liu, Changle and Yu, Rong and Si, Qimiao},
  journal = {Phys. Rev. B},
  volume = {108},
  number = {21},
  pages = {214522},
  year = {2023},
  doi = {10.1103/PhysRevB.108.214522}
}

@article{luo2024high,
  title = {High-{$T_c$} superconductivity in $\mathrm{La}_{3}\mathrm{Ni}_{2}\mathrm{O}_{7}$ based on the bilayer two-orbital t-{J} model},
  author = {Luo, Zhihui and Lv, Biao and Wang, Meng and Wú, Wéi and Yao, Dao-Xin},
  journal = {npj Quantum Mater.},
  volume = {9},
  number = {1},
  pages = {1-7},
  year = {2024},
  doi = {10.1038/s41535-024-00668-w}
}

@article{yang2023possible,
  title = {Possible ${s}_{\ifmmode\pm\else\textpm\fi{}}$-wave superconductivity in $\mathrm{La}_{3}\mathrm{Ni}_{2}\mathrm{O}_{7}$},
  author = {Yang, Qing-Geng and Wang, Da and Wang, Qiang-Hua},
  journal = {Phys. Rev. B},
  volume = {108},
  number = {14},
  pages = {L140505},
  year = {2023},
  doi = {10.1103/PhysRevB.108.L140505}
}

@article{sakakibara2024possible,
  title = {Possible High ${T}_{c}$ Superconductivity in $\mathrm{La}_{3}\mathrm{Ni}_{2}\mathrm{O}_{7}$ under High Pressure through Manifestation of a Nearly Half-Filled Bilayer Hubbard Model},
  author = {Sakakibara, Hirofumi and Kitamine, Naoya and Ochi, Masayuki and Kuroki, Kazuhiko},
  journal = {Phys. Rev. Lett.},
  volume = {132},
  number = {10},
  pages = {106002},
  year = {2024},
  doi = {10.1103/PhysRevLett.132.106002}
}

@article{cao2024flat,
  title = {Flat bands promoted by Hund's rule coupling in the candidate double-layer high-temperature superconductor $\mathrm{La}_{3}\mathrm{Ni}_{2}\mathrm{O}_{7}$ under high pressure},
  author = {Cao, Yingying and Yang, Yi-feng},
  journal = {Phys. Rev. B},
  volume = {109},
  number = {8},
  pages = {L081105},
  year = {2024},
  doi = {10.1103/PhysRevB.109.L081105}
}

@article{yang2023interlayer,
  title = {Interlayer valence bonds and two-component theory for high-${T}_{c}$ superconductivity of $\mathrm{La}_{3}\mathrm{Ni}_{2}\mathrm{O}_{7}$ under pressure},
  author = {Yang, Yi-feng and Zhang, Guang-Ming and Zhang, Fu-Chun},
  journal = {Phys. Rev. B},
  volume = {108},
  number = {20},
  pages = {L201108},
  year = {2023},
  doi = {10.1103/PhysRevB.108.L201108}
}

@article{zhang2023trends,
  title = {Impurity and vortex states in the bilayer high-temperature superconductor $\mathrm{La}_{3}\mathrm{Ni}_{2}\mathrm{O}_{7}$},
  author = {Huang, Junkang and Wang, Z. D. and Zhou, Tao},
  journal = {Phys. Rev. B},
  volume = {108},
  number = {17},
  pages = {174501},
  year = {2023},
  doi = {10.1103/PhysRevB.108.174501}
}

@article{jiang2024pressure,
  title = {Pressure Driven Fractionalization of Ionic Spins Results in Cupratelike High-${T}_{c}$ Superconductivity in $\mathrm{La}_{3}\mathrm{Ni}_{2}\mathrm{O}_{7}$},
  author = {Jiang, Ruoshi and Hou, Jinning and Fan, Zhiyu and Lang, Zi-Jian and Ku, Wei},
  journal = {Phys. Rev. Lett.},
  volume = {132},
  number = {12},
  pages = {126503},
  year = {2024},
  doi = {10.1103/PhysRevLett.132.126503}
}

@article{liu2024electronic,
  title={Electronic correlations and partial gap in the bilayer nickelate $\mathrm{La}_{3}\mathrm{Ni}_{2}\mathrm{O}_{7}$},
  author = {Liu, Zhe and Huo, Mengwu and Li, Jie and Li, Qing and Liu, Yuecong and Dai, Yaomin and Zhou, Xiaoxiang and Hao, Jiahao and Lu, Yi and Wang, Meng and Wen, Hai-Hu},
  journal={Nat. Commun.},
  volume={15},
  number={1},
  pages={7570},
  year={2024},
  publisher={Nature Publishing Group UK London},
  doi = {10.1038/s41467-024-52001-5}
}

@misc{liu2025superconductivitynormalstatetransportcompressively,
    author={Yidi Liu and Eun Kyo Ko and Yaoju Tarn and Lopa Bhatt and Berit H. Goodge and David A. Muller and Srinivas Raghu and Yijun Yu and Harold Y. Hwang},
    year={2025},
    title={Superconductivity and normal-state transport in compressively strained {La$_2$PrNi$_2$O$_7$} thin films},
    eprint={2501.08022},
    archivePrefix={arXiv},
}

@misc{bhatt2025resolvingstructuraloriginssuperconductivity,
      author={Lopa Bhatt and Abigail Y. Jiang and Eun Kyo Ko and Noah Schnitzer and Grace A. Pan and Dan Ferenc Segedin and Yidi Liu and Yijun Yu and Yi-Feng Zhao and Edgar Abarca Morales and Charles M. Brooks and Antia S. Botana and Harold Y. Hwang and Julia A. Mundy and David A. Muller and Berit H. Goodge},
      year={2025},
      title={Resolving Structural Origins for Superconductivity in Strain-Engineered {La$_3$Ni$_2$O$_7$} Thin Films}, 
      eprint={2501.08204},
      archivePrefix={arXiv},
}

@misc{li2025photoemissionevidencemultiorbitalholedoping,
      author={Peng Li and Guangdi Zhou and Wei Lv and Yueying Li and Changming Yue and Haoliang Huang and Lizhi Xu and Jianchang Shen and Yu Miao and Wenhua Song and Zihao Nie and Yaqi Chen and Heng Wang and Weiqiang Chen and Yaobo Huang and Zhen-Hua Chen and Tian Qian and Junhao Lin and Junfeng He and Yu-Jie Sun and Zhuoyu Chen and Qi-Kun Xue},
      year={2025},
      title={Angle-resolved photoemission spectroscopy of superconducting {(La,Pr)}$_3${Ni}$_2${O}$_7$/{SrLaAlO$_4$} heterostructures},
      eprint={2501.09255},
      archivePrefix={arXiv},
}

@article{yue2025correlatedelectronicstructuresunconventional, 
      title={Correlated electronic structures and unconventional superconductivity in bilayer nickelate heterostructures},
   volume={12},
   DOI={10.1093/nsr/nwaf253},
   number={10},
   journal={Natl. Sci. Rev.},
   author={Yue, Changming and Miao, Jian-Jian and Huang, Haoliang and Hua, Yichen and Li, Peng and Li, Yueying and Zhou, Guangdi and Lv, Wei and Yang, Qishuo and Yang, Fan and Sun, Hongyi and Sun, Yu-Jie and Lin, Junhao and Xue, Qi-Kun and Chen, Zhuoyu and Chen, Wei-Qiang},
   year={2025},
   month={June},
   pages = {nwaf253}
}

@article{yang_orbital-dependent_2024,
	title = {Orbital-dependent electron correlation in double-layer nickelate {La$_3$Ni$_2$O$_7$}},
	volume = {15},
	issn = {2041-1723},
	doi = {10.1038/s41467-024-48701-7},
	number = {1},
	urldate = {2025-03-20},
	journal = {Nat. Commun.},
	author = {Yang, Jiangang and Sun, Hualei and Hu, Xunwu and Xie, Yuyang and Miao, Taimin and Luo, Hailan and Chen, Hao and Liang, Bo and Zhu, Wenpei and Qu, Gexing and Chen, Cui-Qun and Huo, Mengwu and Huang, Yaobo and Zhang, Shenjin and Zhang, Fengfeng and Yang, Feng and Wang, Zhimin and Peng, Qinjun and Mao, Hanqing and Liu, Guodong and Xu, Zuyan and Qian, Tian and Yao, Dao-Xin and Wang, Meng and Zhao, Lin and Zhou, X. J.},
	month = may,
	year = {2024},
	pages = {4373},
}

@article{zhu_superconductivity_2024,
	title = {Superconductivity in pressurized trilayer {La$_4$Ni$_3$O$_{10-\delta}$} single crystals},
	volume = {631},
	doi = {10.1038/s41586-024-07553-3},
	number = {8021},
	journal = {Nature},
	author = {Zhu, Yinghao and Peng, Di and Zhang, Enkang and Pan, Bingying and Chen, Xu and Chen, Lixing and Ren, Huifen and Liu, Feiyang and Hao, Yiqing and Li, Nana and Xing, Zhenfang and Lan, Fujun and Han, Jiyuan and Wang, Junjie and Jia, Donghan and Wo, Hongliang and Gu, Yiqing and Gu, Yimeng and Ji, Li and Wang, Wenbin and Gou, Huiyang and Shen, Yao and Ying, Tianping and Chen, Xiaolong and Yang, Wenge and Cao, Huibo and Zheng, Changlin and Zeng, Qiaoshi and Guo, Jian-gang and Zhao, Jun},
	month = jul,
	year = {2024},
	pages = {531--536},
}

@article{ko_signatures_2025,
	title = {Signatures of ambient pressure superconductivity in thin film {La$_3$Ni$_2$O$_7$}},
	volume = {638},
	issn = {1476-4687},
	doi = {10.1038/s41586-024-08525-3},
	number = {8052},
	journal = {Nature},
	author = {Ko, Eun Kyo and Yu, Yijun and Liu, Yidi and Bhatt, Lopa and Li, Jiarui and Thampy, Vivek and Kuo, Cheng-Tai and Wang, Bai Yang and Lee, Yonghun and Lee, Kyuho and Lee, Jun-Sik and Goodge, Berit H. and Muller, David A. and Hwang, Harold Y.},
	month = feb,
	year = {2025},
	pages = {935--940},
}

@article{zhou_ambient-pressure_2025,
	title = {Ambient-pressure superconductivity onset above 40 {K} in ({La},{Pr}){$_3$Ni$_2$O$_7$} films},
	issn = {1476-4687},
	url = {https://www.nature.com/articles/s41586-025-08755-z},
	doi = {10.1038/s41586-025-08755-z},
	journal = {Nature},
        volume={640},
        number={8059},
        pages={641--646},
	author = {Zhou, Guangdi and Lv, Wei and Wang, Heng and Nie, Zihao and Chen, Yaqi and Li, Yueying and Huang, Haoliang and Chen, Weiqiang and Sun, Yujie and Xue, Qi-Kun and Chen, Zhuoyu},
	month = feb,
	year = {2025},

}

@article{wang_pressure-induced_2024,
	title = {Pressure-{Induced} {Superconductivity} {In} {Polycrystalline} {La$_3$Ni$_2$O$_{7-\delta}$}},
	volume = {14},
	doi = {10.1103/PhysRevX.14.011040},
	number = {1},
	urldate = {2025-03-20},
	journal = {Phys. Rev. X},
	author = {Wang, G. and Wang, N.N. and Shen, X. L. and Hou, J. and Ma, L. and Shi, L. F. and Ren, Z. A. and Gu, Y. D. and Ma, H. M. and Yang, P. T. and Liu, Z. Y. and Guo, H. Z. and Sun, J. P. and Zhang, G. M. and Calder, S. and Yan, J.-Q. and Wang, B. S. and Uwatoko, Y. and Cheng, J.-G.},
	month = mar,
	year = {2024},
	pages = {011040},
}

@article{Li_2025,
   title={Identification of Superconductivity in Bilayer Nickelate {La$_3$Ni$_2$O$_7$} under High Pressure up to 100 {GPa}},
   issn={2053-714X},
   volume = {12},
   pages = {nwaf220},
   url={https://doi.org/10.1093/nsr/nwaf220},
   journal={Natl. Sci. Rev.},
   publisher={Oxford University Press (OUP)},
   author={Li, Jingyuan and Peng, Di and Ma, Peiyue and Zhang, Hengyuan and Xing, Zhenfang and Huang, Xing and Huang, Chaoxin and Huo, Mengwu and Hu, Deyuan and Dong, Zixian and Chen, Xiang and Xie, Tao and Dong, Hongliang and Sun, Hualei and Zeng, Qiaoshi and Mao, Ho-kwang and Wang, Meng},
   year={2025},
   month=may,
}

@article{li_signature_2024,
	title = {Signature of {Superconductivity} in {Pressurized} {La$_4$Ni$_3$O$_{10}$}},
	volume = {41},
	issn = {0256-307X},
	doi = {10.1088/0256-307X/41/1/017401},
	number = {1},
	urldate = {2025-03-20},
	journal = {Chin. Phys. Lett.},
	author = {Li, Qing and Zhang, Ying-Jie and Xiang, Zhe-Ning and Zhang, Yuhang and Zhu, Xiyu and Wen, Hai-Hu},
	month = jan,
	year = {2024},
	pages = {017401},
}

@article{PhysRevB.110.134520,
  title = {Tunneling spectra with gaplike features observed in nickelate {{La}}$_{3}${{Ni}}$_{2}${{O}}$_{7}$ at ambient pressure},
  author = {Fan, Shengtai and Luo, Zhihui and Huo, Mengwu and Wang, Zhaohui and Li, Han and Yang, Huan and Wang, Meng and Yao, Dao-Xin and Wen, Hai-Hu},
  journal = {Phys. Rev. B},
  volume = {110},
  issue = {13},
  pages = {134520},
  numpages = {9},
  year = {2024},
  month = {Oct},
  publisher = {American Physical Society},
  doi = {10.1103/PhysRevB.110.134520}
}

@article{ni_spin_2025,
	title = {Spin density wave in the bilayered nickelate {La}$_{3}${Ni}$_{2}${O}$_{7-\delta}$ at ambient pressure},
	volume = {10},
	doi = {10.1038/s41535-025-00740-z},
	number = {1},
	journal = {npj Quantum Mater.},
	author = {Ni, Xiao-Sheng and Ji, Yuyang and He, Lixin and Xie, Tao and Yao, Dao-Xin and Wang, Meng and Cao, Kun},
	month = feb,
	year = {2025},
	pages = {1--9},
}

@article{JM2025,
	title = {Intertwined charge and spin instability of  {La}$_{3}${Ni}$_{2}${O}$_{7}$},
	volume = {68},
	url = {https://doi.org/10.1007/s11433-025-2702-7},
	doi = {10.1007/s11433-025-2702-7},
	journal = {Science China Physics, Mechanics \& Astronomy},
	author = {Jiang, Guiwen and Qin, Chenye and Foyevtsova, Kateryna and Si, Liang and Berciu, Mona and Sawatzky, George A. and Jiang, Mi},
	month = {july},
	year = {2025},
	pages = {297411},
}

@article{PhysRevB.111.075140,
  title = {Origin of the density wave instability in trilayer nickelate {L}{a}$_{4}${N}{{i}}$_{3}${{O}}$_{10}$ revealed by optical and ultrafast spectroscopy},
  author = {Xu, Shuxiang and Chen, Cui-Qun and Huo, Mengwu and Hu, Deyuan and Wang, Hao and Wu, Qiong and Li, Rongsheng and Wu, Dong and Wang, Meng and Yao, Dao-Xin and Dong, Tao and Wang, Nanlin},
  journal = {Phys. Rev. B},
  volume = {111},
  issue = {7},
  pages = {075140},
  numpages = {12},
  year = {2025},
  month = {Feb},
  publisher = {American Physical Society},
  doi = {10.1103/PhysRevB.111.075140},
  url = {https://link.aps.org/doi/10.1103/PhysRevB.111.075140}
}

@article{PhysRevB.110.014503,
  title = {Trilayer multiorbital models of {La}$_{4}${Ni}$_{3}${O}$_{10}$},
  author = {Chen, Cui-Qun and Luo, Zhihui and Wang, Meng and W\'u, W\'ei and Yao, Dao-Xin},
  journal = {Phys. Rev. B},
  volume = {110},
  issue = {1},
  pages = {014503},
  numpages = {11},
  year = {2024},
  month = {Jul},
  publisher = {American Physical Society},
  doi = {10.1103/PhysRevB.110.014503},
  url = {https://link.aps.org/doi/10.1103/PhysRevB.110.014503}
}

@article{PhysRevB.110.235155,
  title = {Correlation effects in a simplified bilayer two-orbital Hubbard model at half filling},
  author = {Yang, Jian-Jian and Yao, Dao-Xin and Wu, Han-Qing},
  journal = {Phys. Rev. B},
  volume = {110},
  issue = {23},
  pages = {235155},
  numpages = {10},
  year = {2024},
  month = {Dec},
  publisher = {American Physical Society},
  doi = {10.1103/PhysRevB.110.235155}
}

@Article{cpl_41_7_077402,
    title = {Normal and Superconducting Properties of {La}$_{3}${Ni}$_{2}${O}$_{7}$},
    journal = {Chin. Phys. Lett.},
    volume = {41},
    number = {7},
    pages = {},
    year = {2024},
    issn = {},
    doi = {10.1088/0256-307X/41/7/077402},	
    url = {http://cpl.iphy.ac.cn/en/article/doi/10.1088/0256-307X/41/7/077402},
    author = {Meng Wang and Hai-Hu Wen and Tao Wu and Dao-Xin Yao and Tao Xiang}
}

@article{NPzhang,
  title = {High-temperature superconductivity with zero resistance and strange-metal behaviour in {La}$_3${Ni}$_2${O}$_{7-\delta}$},
  author = {Zhang, Yanan and Su, Dajun and Huang, Yanen and Shan, Zhaoyang and
Sun, Hualei and 
Huo, Mengwu and
Ye, Kaixin and
Zhang, Jiawen and
Yang, Zihan and
Xu, Yongkang and
Su, Yi and
Li, Rui and
Smidman, Michael and
Wang, Meng and
Jiao, Lin and
Yuan, Huiqiu
},
  journal = {Nat. Phys.},
  volume = {20},
  issue = {7},
  pages = {1269–1273},
  numpages = {12},
  year = {2024},
  doi = {10.1038/s41567-024-02515-y},
  url = {https://doi.org/10.1038/s41567-024-02515-y}
}

@article{Hou_2023,
doi = {10.1088/0256-307X/40/11/117302},
url = {https://dx.doi.org/10.1088/0256-307X/40/11/117302},
year = {2023},
month = {oct},
publisher = {Chinese Physical Society and IOP Publishing Ltd},
volume = {40},
number = {11},
pages = {117302},
author = {Hou, Jun and Yang, Peng-Tao and Liu, Zi-Yi and Li, Jing-Yuan and Shan, Peng-Fei and Ma, Liang and Wang, Gang and Wang, Ning-Ning and Guo, Hai-Zhong and Sun, Jian-Ping and Uwatoko, Yoshiya and Wang, Meng and Zhang, Guang-Ming and Wang, Bo-Sen and Cheng, Jin-Guang},
title = {Emergence of High-Temperature Superconducting Phase in Pressurized {La}$_3${Ni}$_2${O}$_7$ Crystals},
journal = {Chin. Phys. Lett.},
}

@article{PhysRevB.109.144511,
  title = {Theoretical analysis on the possibility of superconductivity in the trilayer Ruddlesden-Popper nickelate {La}$_{4}${Ni}$_{3}${O}$_{10}$ under pressure and its experimental examination: Comparison with {La}$_{3}${{Ni}}$_{2}${{O}}$_{7}$},
  author = {Sakakibara, Hirofumi and Ochi, Masayuki and Nagata, Hibiki and Ueki, Yuta and Sakurai, Hiroya and Matsumoto, Ryo and Terashima, Kensei and Hirose, Keisuke and Ohta, Hiroto and Kato, Masaki and Takano, Yoshihiko and Kuroki, Kazuhiko},
  journal = {Phys. Rev. B},
  volume = {109},
  issue = {14},
  pages = {144511},
  numpages = {10},
  year = {2024},
  month = {Apr},
  publisher = {American Physical Society},
  doi = {10.1103/PhysRevB.109.144511},
  url = {https://link.aps.org/doi/10.1103/PhysRevB.109.144511}
}

@misc{zhou2025originlocalmagneticexchange,
	title={Origin of local magnetic exchange interaction in infiite-layer nickelates}, 
	author={Yanbing Zhou and Dan Zhao and Boyun Zeng and Chengliang Xia and Yu Wang and Hanghui Chen and Tao Wu and Xianhui Chen},
	year={2025},
	eprint={2505.09476},
	archivePrefix={arXiv}
}

@article{Liangle2025,
  title={Angle-resolved photoemission spectroscopy of superconducting {(La, Pr) $_3$Ni$_2$O$_7$/SrLaAlO$_4$} heterostructures},
  author={Li, Peng and Zhou, Guangdi and Lv, Wei and Li, Yueying and Yue, Changming and Huang, Haoliang and Xu, Lizhi and Shen, Jianchang and Miao, Yu and Song, Wenhua and Nie, Zihao and Chen, Yaqi and Wang, Heng and Chen, Weiqiang and Huang, Yaobo and Chen, Zhen-Hua and Qian, Tian and Lin, Junhao and He, Junfeng and Sun, Yu-Jie and Chen, Zhuoyu and Xue, Qi-Kun},
  journal={Natl. Sci. Rev.},
  doi={10.1093/nsr/nwaf205},
  volume = {12},
  number = {10},
  pages = {nwaf205},
  year={2025},
  publisher={Oxford University Press}
}

@article{Zhaoelectronic2025,
  title = {Electronic structure of Ruddlesden-Popper nickelates: Strain to mimic the effects of pressure},
  author = {Zhao, Yi-Feng and Botana, Antia S.},
  journal = {Phys. Rev. B},
  volume = {111},
  issue = {11},
  pages = {115154},
  numpages = {11},
  year = {2025},
  month = {Mar},
  publisher = {American Physical Society},
  doi = {10.1103/PhysRevB.111.115154},
  url = {https://link.aps.org/doi/10.1103/PhysRevB.111.115154}
}

@article{Lienhanced2025,
  title={Enhanced superconductivity in the compressively strained bilayer nickelate thin films by pressure},
  author={Li, Qing and Sun, Jianping and B{\"o}tzel, Steffen and Ou, Mengjun and Xiang, Zhe-Ning and Lechermann, Frank and Wang, Bosen and Wang, Yi and Zhang, Ying-Jie and Cheng, Jinguang and others},
  journal={Nat. Commun.},
  year={2026},
  url={https://www.nature.com/articles/s41467-026-69660-1#citeas},
  volume = {17},
  pages = {3276}
}

@misc{Zhaopressure2026,
      title={Pressure-enhanced superconductivity and its correlation with suppressed resistance dip in {(La,Pr)$_3$Ni$_2$O$_7$} films}, 
      author={Jinyu Zhao and Guangdi Zhou and Shu Cai and Shuaihang Sun and Yaqi Chen and Jing Guo and Yazhou Zhou and Haoliang Huang and Jin-Feng Jia and Yang Ding and Qi Wu and Zhuoyu Chen and Qi-Kun Xue and Liling Sun},
      year={2026},
      eprint={2603.29531},
      archivePrefix={arXiv}
}

@misc{qiu2025interlayer,
      title={Interlayer coupling enhanced superconductivity near 100 {K} in {La$_{3-x}$Nd$_x$Ni$_2$O$_7$}}, 
      author={Zhengyang Qiu and Junfeng Chen and Dmitrii V. Semenok and Qingyi Zhong and Di Zhou and Jingyuan Li and Peiyue Ma and Xing Huang and Mengwu Huo and Tao Xie and Xiang Chen and Ho-kwang Mao and Viktor Struzhkin and Hualei Sun and Meng Wang},
      year={2025},
      eprint={2510.12359},
      archivePrefix={arXiv}
}

@misc{Litheoretical2025,
      title={Theoretical study on the electronic properties and multiorbital models of {La$_3$Ni$_2$O$_7$} thin films on {SrLaAlO$_4$} (001)}, 
      author={Guanlin Li and Cui-Qun Chen and Haoliang Shi and Zhengtao Liu and Hao Ma and Fubo Tian and Dao-Xin Yao and Defang Duan},
      year={2025},
      eprint={2512.17625},
      archivePrefix={arXiv}
}

\end{document}